\pdfoutput=1
\documentclass[aps,arxiv,twocolumn,superscriptaddress]{revtex4-1}
\usepackage{graphicx}
\usepackage{float}
\usepackage{amsmath}
\usepackage{amssymb}
\usepackage{mathtools}
\usepackage[utf8]{inputenc}
\usepackage[T1]{fontenc}
\usepackage[version=4]{mhchem}
\usepackage{booktabs}

\usepackage{microtype}
\usepackage[colorlinks, allcolors=blue]{hyperref}
\hypersetup{pdftitle={Atomic-Scale Properties of SiO2/Pt}, pdfauthor={J.~Qu, A.~Urban}}

\newlength{\colwidth}
\begin{document}

\title{%
  Potential and pH Dependence of the Buried Interface \\
  of Membrane-Coated Electrocatalysts}

\author{Jianzhou Qu}
\author{Alexander Urban}
\email{a.urban@columbia.edu}
\affiliation{%
  Department of Chemical Engineering,
  Columbia University, New York, NY, USA.
}
\affiliation{%
  Columbia Electrochemical Energy Center (CEEC),
  Columbia University, New York, NY, USA.
}
\date{\today}


\begin{abstract}
  Semipermeable silica membranes are attractive as protective coatings
  for metal electrocatalysts such as platinum but their impact on the
  catalytic properties has not been fully understood.
  Here, we develop a first principles formalism to investigate how
  silica membranes interact with the surface of platinum metal
  electrocatalysts to develop a better understanding of the
  membrane-metal interplay.
  By generalizing the concept of Pourbaix diagrams to electrochemical
  solid-solid interfaces, we establish which bonds are formed between
  the \ce{SiO2} membrane and the \ce{Pt(111)} surface in aqueous
  electrolytes for different pH values and potential biases.
  We find that the membrane termination changes as a function of the pH
  and the potential, which affects the adhesion strength and the energy
  requirements for partial membrane detachment, controlling the Pt
  surface area that is accessible for reactant species.
  The charge-transfer between the \ce{Pt} surface and the \ce{SiO2}
  membrane is also pH and potential dependent and results in changes of
  the Pt surface $d$-band states, which are known to correlate with
  catalytic activity.
  Our analysis reveals the complex response of a buried interface to the
  electrochemical environment and identifies trends that are expected to
  apply also to other membrane-coated electrocatalysts.
\end{abstract}

\maketitle

\section{Introduction}
\label{sec:introduction}

Hydrogen production via water electrolysis is an attractive option for the conversion of clean electric energy to synthetic fuels~\cite{am20-2008-2878, arcbe5-2014-429, aem5-2015-1500985, asce7-2019-8006}, but current electrolyzers operate with purified and deionized water to prevent catalyst poisoning or corrosion~\cite{aami9-2017-20585, ne5-2020-367}.
Stabilizing electrocatalysts in wastewater or seawater would remove an important economic barrier for the adoption of this technology.

Encapsulation with semipermeable membranes has recently attracted renewed interest as an inexpensive strategy for the protection of electrocatalysts~\cite{l11-1995-2837, cm13-2001-967, cc0-2003-1522, jpcc111-2007-15133a, ijohe38-2013-13529, ac8-2018-457, ac8-2018-1767a, afm30-2020-1909262}.
Membrane coated metal particles have also been proposed for gas sensor applications~\cite{aami7-2015-3554, aami9-2017-27193}.
Platinum, which is an efficient catalyst for hydrogen evolution reaction (HER)~\cite{jcp26-1957-532, ssp73-75-2000-207}, i.e., the cathodic reaction in electrolyzer cells, can be stabilized in solutions containing cationic species by coating with amorphous silica (\ce{a-SiO2}) membranes~\cite{ac8-2018-457, ac8-2018-1767a, ne5-2020-367}.
Such \ce{SiO2}-encapsulated Pt catalysts have been extensively characterized, and it has been demonstrated that ultrathin coatings do not lead to notable transport limitations and improve not only the stability but also the catalytic selectivity~\cite{nl16-2016-6452, ac8-2018-1767a, ac8-2018-11423}.

Cyclic voltammetry measurements show that the HER reaction mechanism at the buried \ce{SiO2 / Pt} interface of encapsulated Pt catalysts is different from the mechanism observed for the bare Pt surface in direct contact with water~\cite{jes99-1952-169, ac8-2018-1767a}.
Hence, the \ce{SiO2} membrane is not an inactive bystander that only protects the active phase but actively participates in the catalytic reaction.
This synergy of metal surface and coating is an opportunity for the discovery of efficient and stable catalysts based on more earth-abundant metal species~\cite{acag220-2001-79, csr44-2015-5148}, but a better understanding of the interplay of metal and coating is needed to enable the targeted design of synergistic membrane-coated electrocatalysts in which Pt is replaced with other metals.

Owing to the structural complexity and narrow width of the buried catalyst/coating interface, it is challenging to investigate its atomic-scale properties experimentally~\cite{coissams10-2006-26}.
First principles atomistic calculations are commonly used as a complementary characterization technique for electrocatalysts~\cite{pnas108-2011-937}, and the mechanism of HER over transition-metal surfaces has been extensively investigated with density-functional theory (DFT)~\cite{pccp9-2007-3241, ss606-2012-679, ss640-2015-36}.
Various oxide-supported metal catalysts have been previously modeled with DFT~\cite{jcp133-2010-164703, ac7-2017-4707}, which also includes an investigation of Pt nanoparticles supported on crystalline silica, showing that binding with siloxide groups reduce the electron density of Pt surfaces~\cite{jpcc119-2015-19934, jpcc123-2019-12706}.
Graphene-coated confined metallic electrocatalysts have been studied both experimentally and theoretically~\cite{nl16-2016-6058, acie56-2017-12883, pnas114-2017-5930}, though the interaction between graphene and metal surfaces is mostly limited to weak van der Waals bonding.
Lithium-permeable coatings on electrodes in lithium ion batteries have been investigated using atomistic modeling~\cite{nrm5-2020-105, esm29-2020-71}, although most DFT studies of complex solid-solid interfaces have been limited to interface properties that can be derived from independent calculations of the two solids in contact~\cite{cm28-2016-266, jomca4-2016-3253, j3-2019-1252, nrm5-2020-105}.

In the present work, we employ DFT calculations of buried \ce{SiO2 / Pt} interface structures to determine how \ce{SiO2} membranes attach to the surface of \ce{Pt} catalysts at different electrocatalytic conditions.
We consider a crystalline model interface structure to facilitate exploring different chemical bonds and interface compositions.
In the spirit of Pourbaix diagrams~\cite{jes111-1964-14C}, we introduce a new kind of interface phase diagram (or \emph{interface Pourbaix diagram}) that maps the bonding between the semipermeable membrane and the metal surface as a function of the pH value and the applied bias potential.
We discuss implications for the electrochemical properties of the catalyst and steric effects that may alter the reaction mechanism at the buried interface.

In the following \emph{methods section}, computational details are given and the interface Pourbaix diagram formalism is developed.
The formalism is then applied to the buried \ce{SiO2 / Pt} interface, and results of the findings are reported in the \emph{results section} along with a computational characterization of the stable interface compositions.
In the \emph{discussion section}, the observations are interpreted and compared to experimental data from the literature.

\section{Methods}
\label{sec:methods}

\subsection{Details of the density-functional theory calculations}
\label{sec:DFT}

All density-functional theory (DFT) calculations were performed using the \emph{Vienna Ab Initio Simulation Package} (VASP)~\cite{prb47-1993-558,prb49-1994-14251, cms6-1996-15, prb54-1996-11169} and projector-augmented wave (PAW)~\cite{prb50-1994-17953, prb59-1999-1758} pseudopotentials.
The exchange-correlation functional by Perdew, Burke, and Ernzerhof (PBE)~\cite{prl77-1996-3865, prl78-1997-1396} was employed.
The plane-wave cutoff energy was 520~eV, and the convergence threshold for the self-consistent energy was chosen to be $10^{-5}$~eV.
The atomic forces in optimized structures were less than 0.01~eV/\AA{}.
For the integration of the Brillouin zone, automatic Gamma-centered k-point meshes were used with
$N_{i}= \lfloor{}\max(1, 25 |\vec{b}_{i}|)\rfloor{}$ points in
reciprocal lattice direction $i$, where $|\vec{b}_{i}|$ is the length of the $i$-th reciprocal lattice vector.
For the \ce{SiO2 / Pt} interface structure models, this k-point spacing corresponds to $5\times{}5\times{}1$ k-point meshes.

To capture weak van-der-Waals interactions between the \ce{SiO2} membrane and the Pt surface, we employed Grimme's D3 correction with Becke-Jonson damping~\cite{tjocp132-2010-154104, jocc32-2011-1456}.

For the calculation of the density of states (DOS), k-point meshes with twice finer resolution ($10\times{}10\times{}2$) were employed to ensure convergence.
Additionally, a Gaussian broadening with standard deviation 0.1~eV was applied.

\subsection{Atomic structure}
\label{sec:structures}

The DFT lattice parameters of both face-centered cubic \ce{Pt} and \ce{SiO2} in the $\alpha$-quartz crystal structure are in good agreement with experimental reference values (see \textbf{Table~S1} in the Supporting Information).

Based on the optimized Pt crystal structure, a slab model of the \ce{Pt(111)} surface with $2\times{}2$ surface unit cell and 6~layers was constructed.
The positions of the atoms in the topmost three layers were optimized, while the lowest three layers were kept fixed on their bulk positions in all calculations.

\begin{figure*}
  \centering
  \includegraphics[width=\textwidth]{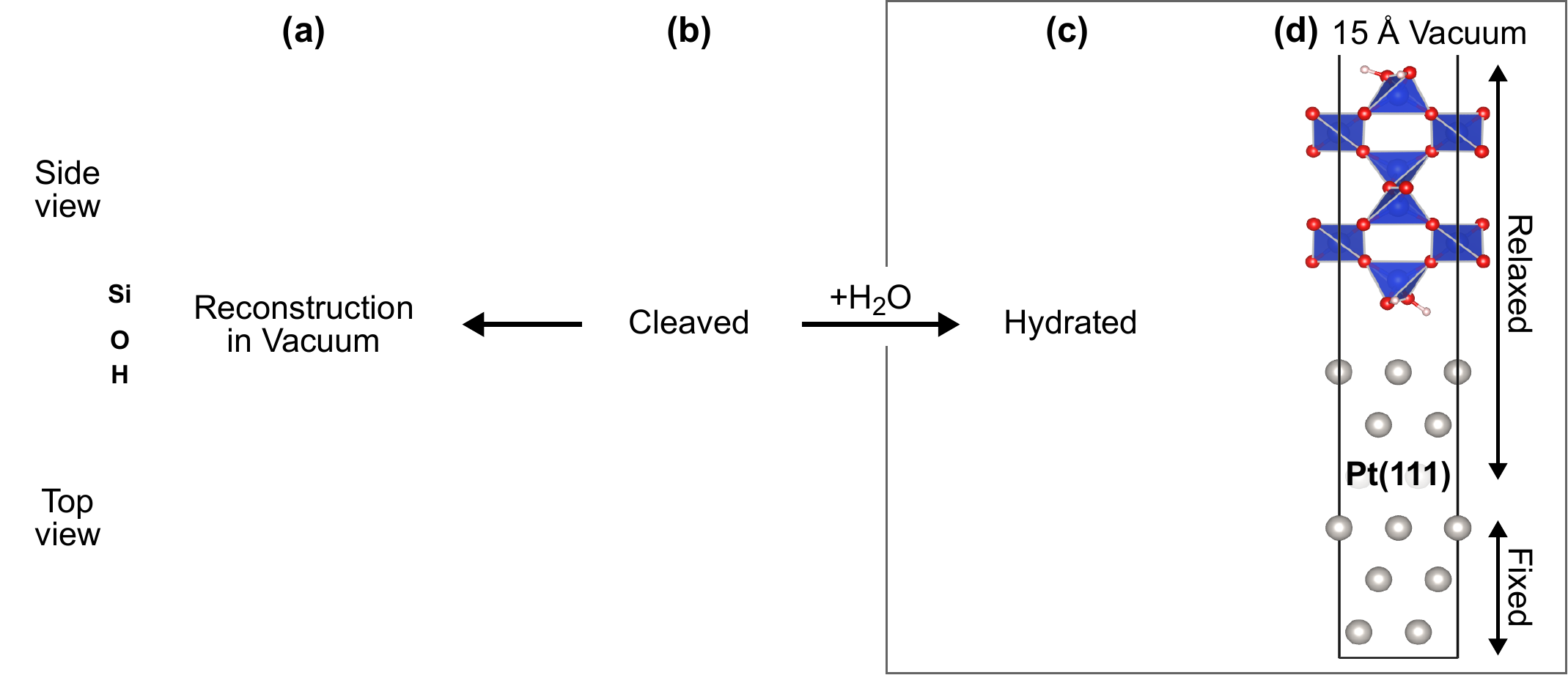}
  \caption{Side and top views of the structures of the \ce{SiO2(001)} surface \textbf{(b)}~as cleaved from the bulk crystal structure, \textbf{(a)}~after reconstruction in vacuum, and \textbf{(c)}~after addition of a water molecule (hydration).  Oxygen is shown as red balls, Si blue, and hydrogen white.  \textbf{(d)}~A representative slab model of the \ce{SiO2 / Pt} interface with hydrated \ce{SiO2} membrane.}
  \label{fig:SiO2-reconstruction}
\end{figure*}

The structure of the membrane was based on the $\alpha$-quartz crystal structure of \ce{SiO2}, since it is the most stable polymorph of \ce{SiO2}~\cite{sr5-2015-14545}.
A slab model of the \ce{SiO2(001)} surface with 6~\ce{SiO2}~layers was constructed and the positions of all atoms were optimized.
The \ce{SiO2(001)} surface is unstable as cleaved from the bulk crystal structure (\textbf{Figure~\ref{fig:SiO2-reconstruction}b}) and reconstructs in vacuum (\textbf{Figure~\ref{fig:SiO2-reconstruction}a})~\cite{apl93-2008-181911}.
In the presence of water, the \ce{SiO2(001)} surface forms a hydrated reconstruction through the addition of one water molecule per surface unit cell (\textbf{Figure~\ref{fig:SiO2-reconstruction}c})~\cite{pccp9-2007-2146}.
We considered the exposed \ce{SiO2} surface to be generally hydrated, since the coated electrode is submerged in aqueous electrolyte at electrocatalytic conditions.

Our \ce{SiO2 / Pt} interface structure model was constructed by combining the \ce{Pt(111)} and \ce{SiO2(001)} slabs  (\textbf{Figure~\ref{fig:SiO2-reconstruction}d}).
Both the \ce{Pt(111)} surface and the \ce{SiO2(001)} surface have hexagonal unit cells, so that an interface structure could be constructed by scaling the $a$ and $b$ lattice parameters of the \ce{SiO2(001)} slab model to the dimensions of the \ce{Pt(111)} surface unit cell.
The lattice mismatch of $5.63$~\AA{} (\ce{Pt}) to $5.02$~\AA{} (\ce{SiO2}) corresponds to 12\% tensile strain in the \ce{SiO2} membrane.
The optimal relative position of the Pt surface and the membrane was first pre-screened using a binding site scan (\textbf{Figure~S1}), which was followed by an optimization of the positions of all atoms in the \ce{SiO2} membrane and the Pt surface, as described in section~\ref{sec:DFT}.
The interface slab model was terminated by a 15~\AA{} wide vacuum region in the direction of the exposed \ce{SiO2(001)} surface.
A representative slab model with hydrated membrane is shown in \textbf{Figure~\ref{fig:SiO2-reconstruction}d}, and structure files are provided as part of the \textbf{Supporting Information}.

\subsection{Interface and surface stability}
\label{sec:interface-energy}

We introduce three quantities related to different aspects of the stability of interfaces: (i)~the \emph{adhesion energy}, (ii)~the \emph{detachment energy}, and (iii)~the \emph{interface free energy of formation}.

The interface \emph{adhesion energy} originates from attractive interactions between the Pt surface and the \ce{SiO2} membrane
\begin{align}
  E_{\textup{ad}}
  = \frac{1}{A}  \Bigl(
    E_{\ce{SiO2 / Pt}} -  E_{\ce{Pt}} - E_{\ce{SiO2}}
  \Bigr)
  \quad ,
  \label{eq:ahesion-energy}
\end{align}
where $E_{\ce{SiO2 / Pt}}$ is the energy of the interface structure as predicted by DFT and $E_{\ce{Pt}}$ and $E_{\ce{SiO2}}$ are the DFT energies of the \ce{Pt(111)} and \ce{SiO2} slabs in isolation.

The adhesion energy needs to be overcome to fully detach the \ce{SiO2} membrane from the Pt surface.
For reactants to reach the Pt catalyst surface, it is sufficient if the membrane detaches locally from the Pt surface by a few Angstroms depending on the size of the reactant species.
The energy required for such partial detachment of the membrane is more intuitively expressed relative to the energy of the intact interface.
We therefore define the \emph{detachment energy} as the energy difference between the interface ground-state structure and a structure in which the membrane has dissociated by a distance $r$ from the Pt surface
\begin{align}
  E_{\textup{detach}}
  = \frac{1}{A}  \Bigl(
    E_{\ce{SiO2 / Pt}}^{\textup{detach}}(r) -  E_{\ce{SiO2 / Pt}}
  \Bigr)
  \quad ,
  \label{eq:detachment-energy}
\end{align}
where $E_{\ce{SiO2 / Pt}}^{\textup{detach}}$ and $E_{\ce{SiO2 / Pt}}$ are the DFT energies of the detached and the ground-state interface structures, respectively, and $A$ is the cross area of the interface.

Although the adhesion energy may be considered a measure of mechanical
stability, it does not quantify the thermodynamic stability of the interface.
Instead, the \emph{interface free energy of formation} has to be considered, which is a function of other thermodynamic variables and is therefore environment-dependent.
An approximation to the free energy of formation is derived in the following section.

\subsection{Interface Pourbaix diagram formalism}
\label{sec:interface-formation-energy}

In an electrochemical cell, the encapsulated electrode is submerged in an electrolyte and a potential bias may be applied.
Assuming that the \ce{SiO2} membrane is permeable for water and protons~\cite{jopcm28-2015-23001, aami11-2019-43130, afm30-2020-1909262}, in thermodynamic equilibrium, the buried interface is subject to the pH value of the aqueous electrolyte and the applied potential.
At the most fundamental level, the thermodynamic stability of a given interface structure is therefore determined by the pH- and potential-dependent free energies of all competing interface structures.

The stability of phases in water as function of the pH value and potential is conventionally characterized by Pourbaix diagrams~\cite{Pourbaix1974}.
Methodologies for calculating Pourbaix diagrams by combining results from DFT and experiment have previously been described~\cite{prb85-2012-235438a, jpcl6-2015-1785a, pccp18-2016-29561}.
Here, we generalize the concept of computational Pourbaix diagrams to interfaces of (semi-)permeable membranes.

To compare the relative stability of different interface structures with varying oxygen and hydrogen contents at the buried \ce{SiO2 / Pt} interface, we consider the free energy of reaction of the formal interface formation reaction
\begin{widetext}
\begin{align}
  \ce{
    Pt(111)-slab + SiO2(001)-slab + a H2O + b (H^{+} + e^{-})
    <=> SiO_{2} * a H2O * b H / Pt
  }
  \; ,
\end{align}
\end{widetext}
where the $a$ and $b$ are the number of water molecules and proton-electron pairs added (or removed) during the formation of the interface structure.
For our specific \ce{SiO2} slab model with 6~\ce{SiO2} layers (see section~\ref{sec:structures}), and assuming that the \ce{SiO2} surface that is exposed to the aqueous electrolyte is always hydrated, the composition of the \ce{SiO2} slab can be explicitly written as \ce{6 SiO2 * H2O} = \ce{Si6H2O13}, yielding
\begin{widetext}
\begin{align}
  \ce{Pt + Si6H2O13 + $(y-13)$ H2O + $(x-2y+24)$ H^{+}
      + $(x-2y+24)$e^{-} <=> Si6H_{x}O_{y} / Pt}
  \label{eq:interface-formation-reaction}
  \; ,
\end{align}
\end{widetext}
where Pt is used as a shorthand for the \ce{Pt(111)} slab.

Two assumptions are made by casting all interface compositions \ce{Si6H_{x}O_{y} / Pt} in the form of equation~\eqref{eq:interface-formation-reaction}.
First, we consider only interfaces that can be constructed from the Pt slab and \ce{SiO2} slab models described in section~\ref{sec:structures}.
Using different slab models (e.g., with fewer numbers of layers or larger surface unit cells) as reference would change the free energy of reaction~\eqref{eq:interface-formation-reaction} by a constant but would not affect relative free energies and relative interface stability.
Second, since we are interested only in the nature of the buried interface, we consider the exposed \ce{SiO2} surface to be always hydrated, and only variation of H and O atoms at the buried interface is considered.
This means, we assume that our membrane structure model is thick enough, so that the termination of the exposed \ce{SiO2(001)} surface can be expected to be independent of the composition at the buried interface and would therefore not affect the relative stability of different interface compositions.

With reaction~\eqref{eq:interface-formation-reaction}, the relative Gibbs free energy of formation of a specific interface with composition \ce{Si6H_{x}O_{y} / Pt} is
\begin{align}
\begin{aligned}
  \Delta G_{\textup{i}}
  &= G_{\ce{Si6H_{x}O_{y} / Pt}} - G_{\ce{Pt}} - G_{\ce{Si6H2O13}} \\
  &- (y-13)\, G_{\ce{H2O}}
   - (x-2y+24)\, \bigl(\mu_{\ce{H^+}} + \mu_{\ce{e^-}}\bigr)
\end{aligned}
\label{eq:interface-formation-energy}
\end{align}
where $G_{A}$ denotes the Gibbs free energy of phase $A$, and $\mu_{\ce{H^+}}$ and $\mu_{\ce{e^-}}$ are the chemical potentials of protons and electrons, respectively.
We note that proton exchange membrane (PEM) electrolyzers operate in acidic conditions (pH~2),\cite{ijhe38-2013-4901} the pH value in alkaline electrolyzers is acidic (pH~1–4) at the anode and alkaline (pH~14) at the cathode,\cite{pieee100-2012-410} and the pH range for saline water electrolysis is between pH 4–9.\cite{ne5-2020-367}
Since the water self-ionization can be neglected over the relevant pH range, the concentration of hydroxide ions (\ce{OH^-}) is inversely proportional to the \ce{H^+} concentration (pOH~$\approx{}14-$~pH).
Therefore, \ce{OH^-} ions do not occur explicitly in equation~\eqref{eq:interface-formation-energy} irrespective of the pH value of the electrolyte.

At room temperature, the temperature dependence of the free energy of solid and liquid phases is small in comparison to the dependence on the proton and electron chemical potentials (see section~3 of the supporting information), so that we approximate the free energies of the interface structure, the \ce{Pt} slab, the \ce{SiO2} slab, and water with their respective 0~K enthalpies as approximated by DFT energies:
\begin{align}
\begin{aligned}
  G_{\ce{Si6H_{x}O_{y} / Pt}} &\approx E_{\ce{Si6H_{x}O_{y} / Pt}} \\
  G_{\ce{Pt(111)}} &\approx E_{\ce{Pt(111)}} \\
  G_{\ce{Si6H2O13}} &\approx E_{\ce{Si6H2O13}}
  \quad .
\end{aligned}
\label{eq:zero-Kelvin-approximation}
\end{align}
We recall also that the energies of the \ce{Pt} slab and \ce{Si6H2O13} are constant terms that do not affect the relative free energies.

The chemical potentials of protons and electrons can be expressed in terms of the pH value and the applied potential $U$, respectively
\begin{align}
  \mu_{\ce{H^+}} = \mu_{\ce{H^+}}^{\circ} - 2.3 k_{\textsc{b}} T \textup{pH}
  \quad\text{and}\quad
  \mu_{\ce{e^-}} = \mu_{\ce{e^-}}^{\circ} - e U
  \label{eq:mu_H+-and-mu_e-}
\end{align}
where $k_{\textsc{b}}$ is Boltzmann's constant, and $T$ is the temperature.
At standard conditions and in equilibrium, the chemical potentials are, further, related to the chemical potential of hydrogen in the gas phase (\emph{standard hydrogen electrode})
\begin{align}
  \frac{1}{2} \mu_{\ce{H2}}^{\circ}
  = \mu_{\ce{H^+}}^{\circ} + \mu_{\ce{e^-}}^{\circ}
  \quad\text{with}\quad
  \mu_{\ce{H2}}^{\circ} = H_{\ce{H2}}^{\circ} - T S_{\ce{H2}}^{\circ}
  \label{eq:SHE}
  \quad .
\end{align}
The measured entropy of hydrogen at standard conditions can be taken from thermochemical tables~\cite{Chase-1998-NIST} and is $S_{\ce{H2}}^{\circ}=130.680$~J\,K$^{-1}$\,mol$^{-1}= 0.00135$~eV\,K$^{-1}$.
To achieve compatibility with our DFT calculations, the enthalpy of hydrogen at standard conditions is approximated by the DFT energy of a hydrogen molecule $H_{\ce{H2}}^{\circ}\approx{}E_{\ce{H2}}$, which yields
\begin{align}
  \mu_{\ce{H^+}}^{\circ} + \mu_{\ce{e^-}}^{\circ}
  \approx \frac{1}{2} \Bigl( E_{\ce{H2}} - T S_{\ce{H2}}^{\circ} \Bigr)
  \label{eq:CSHE}
\end{align}
which is also known as the \emph{computational standard hydrogen
  electrode}~\cite{jpcb108-2004-17886}.

Inserting the approximation of equations
\eqref{eq:zero-Kelvin-approximation}~and~\eqref{eq:CSHE} into the
expression of the free energy of formation in
equation~\eqref{eq:interface-formation-energy} gives
\begin{align}
  &\Delta G_{\textup{interface}}(\textup{pH}, U)
  \label{eq:interface-formation-energy-approx} \\\nonumber
  &\approx E_{\ce{Si6H_{x}O_{y} / Pt}} - E_{\ce{Pt}}
   - E_{\ce{Si6H2O13}} - (y-13)\, E_{\ce{H2O}}\\\nonumber
  &- (x-2y+24)\, \Bigl[
      \frac{1}{2} \Bigl(E_{\ce{H2}} - T S_{\ce{H2}}^{\circ}\Bigr)
  \quad - 2.3 k_{\textsc{b}} T \textup{pH}
         - e U
    \Bigr]
\end{align}
which depends only on DFT energies and the tabulated hydrogen entropy.

Using the expression of the interface free energy of formation in equation~\eqref{eq:interface-formation-energy-approx}, interface Pourbaix diagrams can be constructed by determining the lower convex hull of the free energy functions of all relevant interface structures.

\begin{figure*}
  \centering
  \includegraphics[width=\textwidth]{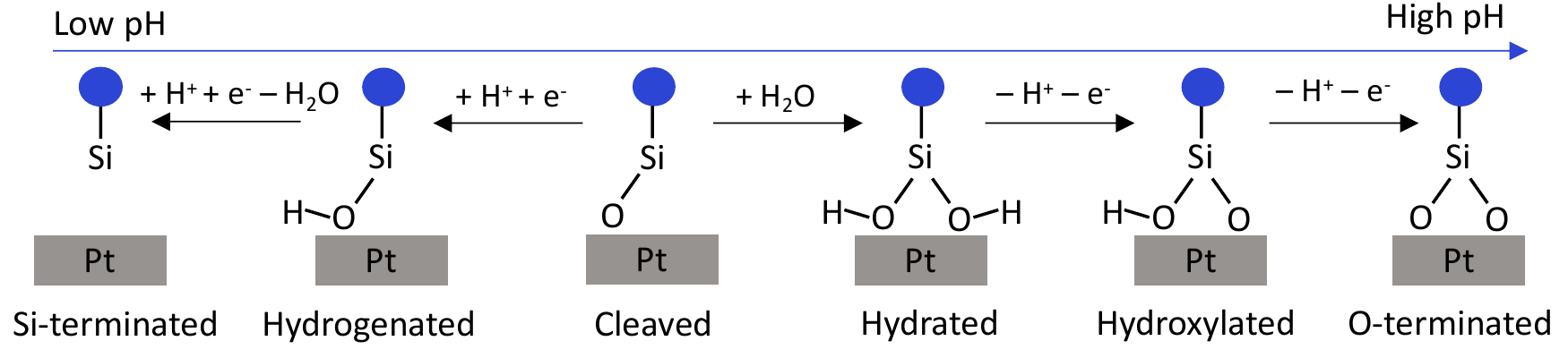}
  \caption{Schematic of the considered \ce{SiO2} membrane terminations in contact with the Pt surface at the buried interface.  The different membrane terminations are interrelated by the addition and removal of protons and water molecules. }
  \label{fig:bond-schematic}
\end{figure*}

\begin{table*}
  \centering
  \caption{Interface adhesion energies and separation of the \ce{SiO2} membrane from the Pt surface for the different membrane terminations shown in \textbf{Figure~\ref{fig:adhesion-energies}}.}
  \label{tbl:adhesion-energies}
  \begin{tabular}{lccc}
    \toprule
    \textbf{Termination} & \textbf{Composition}
    & \textbf{Adhesion Energy (eV\,\AA{}$^{-2}$)} & \textbf{Distance (\AA{})} \\
    \midrule
    Si-terminated & \ce{Si6H2O12 / Pt} & $-0.146$ & 1.82 \\
    Hydrogenated  & \ce{Si6H3O13 / Pt} & $-0.145$ & 2.24 \\
    Cleaved       & \ce{Si6H2O13 / Pt} & $-0.106$ & 1.97 \\
    Reconstructed & \ce{Si6H2O13 / Pt} & $-0.027$ & 2.83 \\
    Hydrated      & \ce{Si6H4O14 / Pt} & $-0.026$ & 2.37 \\
    Hydroxylated  & \ce{Si6H3O14 / Pt} & $-0.121$ & 2.10 \\
    O-terminated  & \ce{Si6H2O14 / Pt} & $-0.159$ & 1.91 \\
    \bottomrule
  \end{tabular}
\end{table*}

\section{Results}
\label{sec:results}

Having developed both structure models (section~\ref{sec:structures}) and a formalism for the prediction of interface stability (section~\ref{sec:interface-formation-energy}), we proceed to investigate the properties of the buried \ce{SiO2 / Pt} interface.

\subsection{Binding of the \texorpdfstring{\ce{SiO2}}{SiO2} membrane to the Pt surface}
\label{sec:binding}

From experimental characterization it is known that the composition of silica membranes can deviate significantly from the ideal \ce{SiO2} and can exhibit different Si:O ratios at the buried interface~\cite{l11-1995-1049}.
To investigate how the chemical termination of the membrane affects the interactions with the \ce{Pt} surface, the termination of the \ce{SiO2} membrane at the interface was systematically varied based on the slab model of \textbf{Figure~\ref{fig:SiO2-reconstruction}d}.

\begin{figure*}
  \centering
  \includegraphics[width=0.8\textwidth]{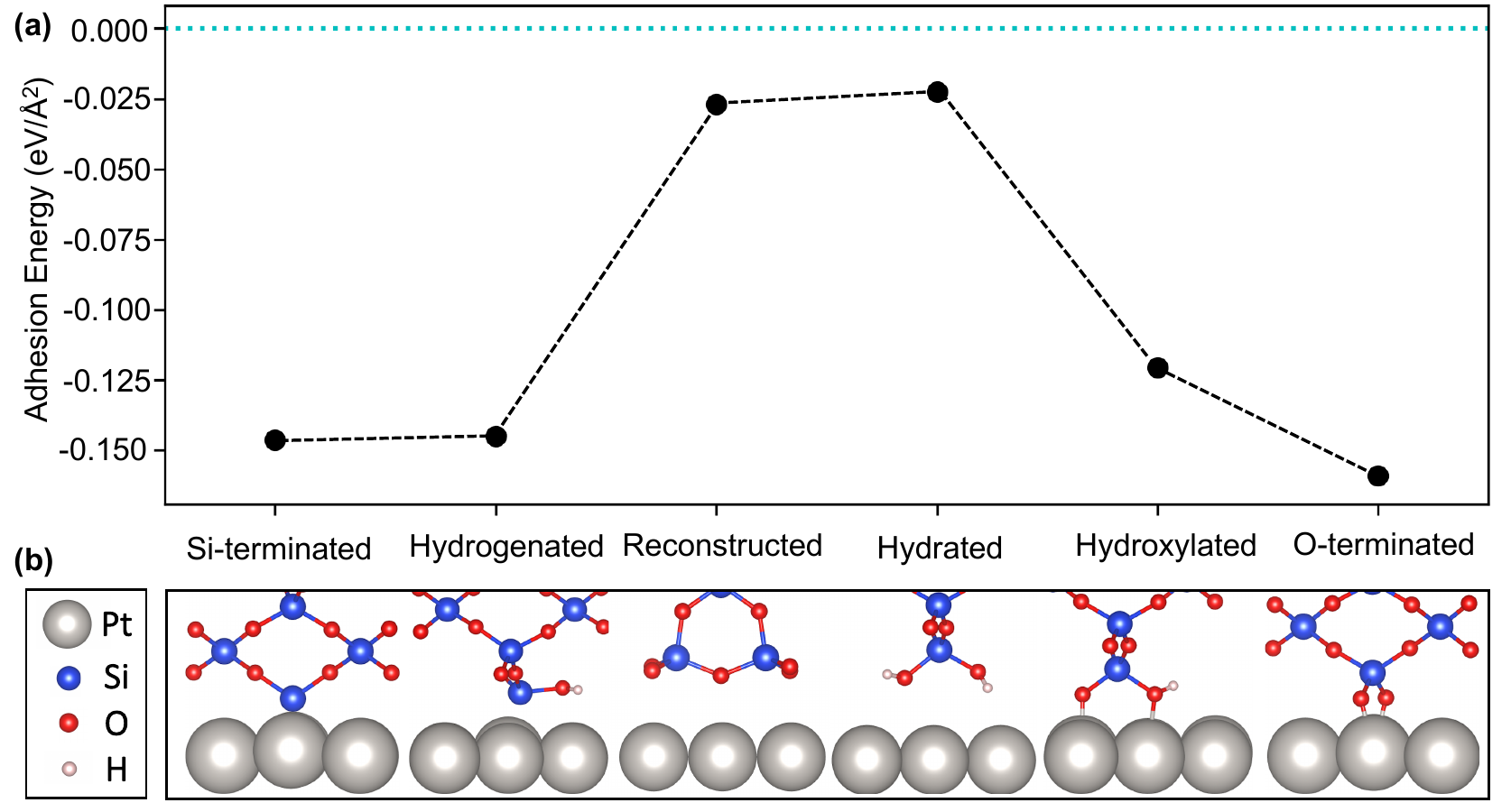}
  \caption{\textbf{(a)}~Adhesion energy of \ce{SiO2 / Pt} interfaces with varying membrane terminations calculated according to equation~\eqref{eq:ahesion-energy}.  The black points indicate the adhesion energies of the different chemistries.  \textbf{(b)}~The atomic structures in the interface regions.  All interface structures were based on the interface model of \textbf{Figure~\ref{fig:SiO2-reconstruction}(d)}.  The compositions of the different interface structures are listed in \textbf{Table~\ref{tbl:adhesion-energies}}.}
  \label{fig:adhesion-energies}
\end{figure*}

By design, our interface structure model exhibits exactly one \ce{SiO4} group per surface unit cell that is in contact with the \ce{Pt} surface.
\textbf{Figure~\ref{fig:bond-schematic}} shows the different membrane terminations that were derived from the ideal cleaved \ce{SiO2} surface and are interrelated by the addition and removal of protons and water molecules.
We also considered the reconstruction of the cleaved membrane surface described in \textbf{Figure~\ref{fig:SiO2-reconstruction}}, which is referred to as \emph{reconstructed} \ce{SiO2} membrane in the following.

The membrane terminations in \textbf{Figure~\ref{fig:bond-schematic}} are sorted in the order of the expected pH dependence based on the number of protons that were formally added or removed to the structures.
\emph{Hydrogenation} of the O atom of the cleaved \ce{SiO2} surface creates a silanol (\ce{Si-OH}) group.
Reaction with an additional proton and the removal of an \ce{H2O} molecule yields a \emph{Si-terminated} membrane.
A \emph{hydrated} membrane is obtained by reaction with a water molecule, as shown also in \textbf{Figure~\ref{fig:SiO2-reconstruction}c}.
Removal of a proton from the hydrated membrane gives a \emph{hydroxylated} termination, since it is related to the cleaved \ce{SiO2} structure by the formal addition of a hydroxy group.
Formally, the simultaneous addition of a water molecule and removal of a proton is equivalent to the addition of a hydroxy group at high pH values.
Removal of the remaining proton from the hydroxylated membrane results in a fully \emph{O-terminated} membrane.

The optimized atomic structures of the different interface terminations along with their respective adhesion energies are shown in \textbf{Figure~\ref{fig:adhesion-energies}}, and the numeric values of the adhesion energies are listed in \textbf{Table~\ref{tbl:adhesion-energies}}.
As seen in \textbf{Figure~\ref{fig:adhesion-energies}a}, the adhesion energy varies with the interface composition over a range of $-0.15$~to~$0.03$~eV\,\AA{}$^{-2}$.
Excerpts from the corresponding atomic structures at the buried interface are seen in \textbf{Figure~\ref{fig:adhesion-energies}b}, and visualizations of the complete slab model structures are shown in supporting \textbf{Figures~S2 and~S3}.
The reconstructed and hydrated \ce{SiO2} membrane terminations show only weak interactions with the \ce{Pt} surface, whereas the Si-, O-terminated, and hydrogenated membranes binds strongly with the \ce{Pt} surface.
The magnitude of the adhesion energy decreases in the order O-terminated $>$ Si-terminated $>$ hydrogenated $>$ hydroxylated $>$ reconstructed $>$ hydrated.

As seen in supporting \textbf{Figure~S4}, the trends are also observed in DFT calculations without van-der-Waals correction, though the adhesion energy of the reconstructed and hydrated membranes with the Pt surface is nearly zero, indicating that these membranes bind mainly via dispersive interactions, and a van-der-Waals correction is indeed needed.

The differences in adhesion energy are also mirrored by variations in the distance between the \ce{Pt} surface and the \ce{SiO2} membrane.
The weakly binding reconstructed and hydrated interfaces show large separations of $\sim$2.8~\AA{} and $\sim$2.4~\AA{} from the Pt surface, respectively, whereas the distances of the strongly binding Si-terminated and O-terminated membranes are only 1.8~\AA{} and 1.9~\AA{}, respectively.
The bond distances found for all interface terminations are listed in \textbf{Table~\ref{tbl:adhesion-energies}}.

A strong binding between the \ce{SiO2} membrane and the \ce{Pt} surface does not necessarily imply that an interface termination is also thermodynamically stable.
As discussed in section~\ref{sec:interface-formation-energy}, the interface stability is determined by the environment-dependent Gibbs free energy of formation, which we will consider in the following section.

\subsection{pH and potential dependence}
\label{sec:pH-and-potential-dependence}

\begin{table*}
  \centering
  \caption{Expressions of the Gibbs formation energies of the different \ce{SiO2 / Pt} interface terminations in \textbf{Figure~\ref{fig:adhesion-energies}} as function of the pH value and the applied potential $U$ (in V vs.~\ce{H/H^+}).  The composition of each interface termination is expressed relative to the composition of the reconstructed membrane.  The Pt substrate is omitted in the reactions for clarity.}
  \label{tbl:interface-free-energies}
  \small
  \begin{tabular}{lll}
    \toprule
    \textbf{Termination} & \textbf{Composition} & $\mathbf{\Delta G_{\textup{interface}}}$ (eV)\\
    \midrule
    Si-terminated & \ce{Si6H2O12 <=> Si6H2O13 - H2O + 2(H^+ + e^-)} & $-0.21 + 0.118\, \textup{pH} + 2\, U$ \\
    Hydrogenated  & \ce{Si6H3O13 <=> Si6H2O13 + 1(H^+ + e^-)} & $-0.04 + 0.059\, \textup{pH} + 1\, U$ \\
    Reconstructed & \ce{Si6H2O13 <=> Si6H2O13} & $0.00$ \\
    Hydrated      & \ce{Si6H4O14 <=> Si6H2O13+H2O} & $0.44$ \\
    Hydroxylated  & \ce{Si6H3O14 <=> Si6H2O13 + H2O - 1(H^+ + e^-)} & $1.14 - 0.059\, \textup{pH} - 1\, U$ \\
    O-terminated  & \ce{Si6H2O14 <=> Si6H2O13 + H2O - 2(H^+ + e^-)} & $2.25 - 0.118\, \textup{pH} - 2\, U$ \\
    \bottomrule
  \end{tabular}
  \vspace{-0.5\baselineskip}
\end{table*}

\begin{figure*}
  \centering
  \includegraphics[width=\textwidth]{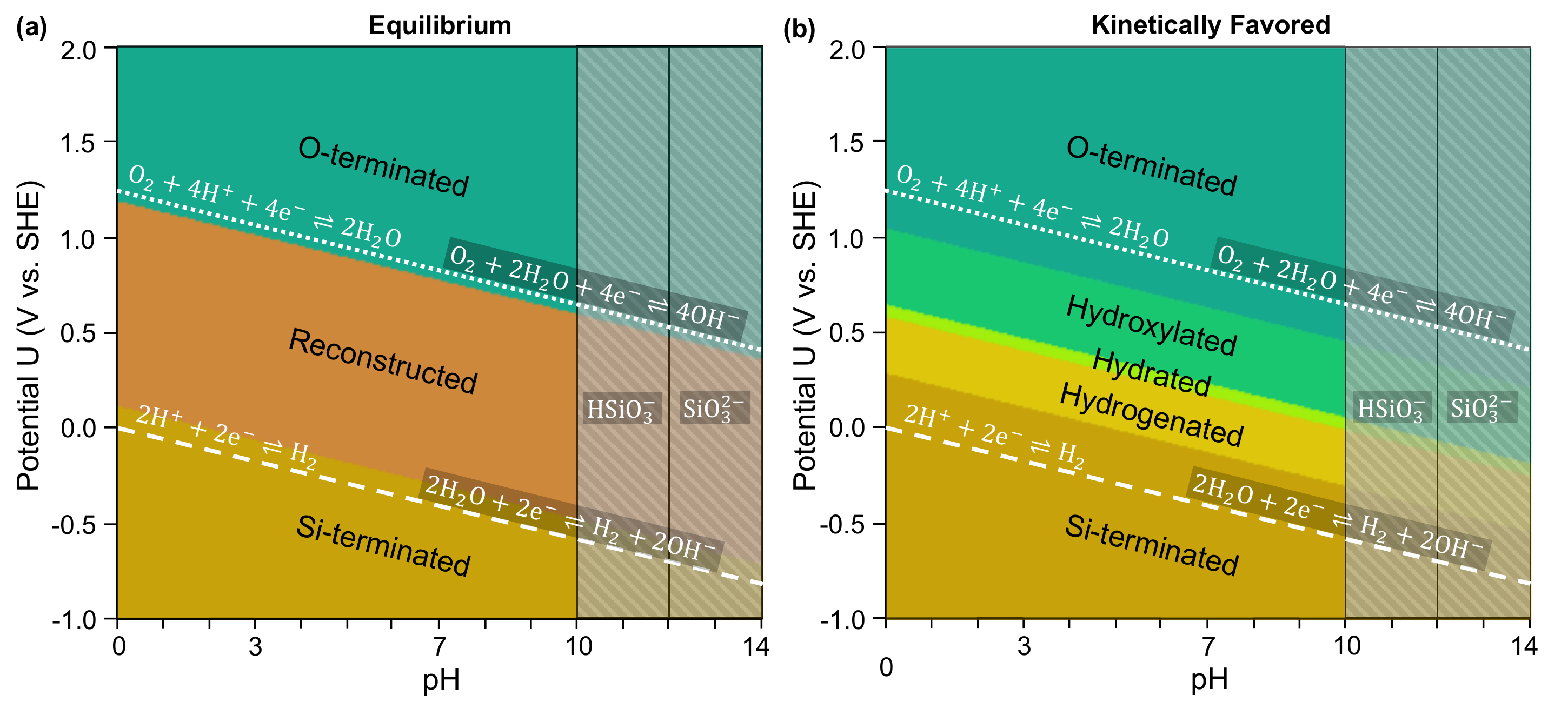}
  \caption{Calculated Pourbaix diagrams of the \ce{SiO2 / Pt} interface \textbf{(a)}~assuming no transport limitation for water through the \ce{SiO2} membrane and \textbf{(b)}~with kinetically stabilized regions due to transport limitations.  Colors indicate stability regions of different interface terminations.  The notation of \textbf{Figure~\ref{fig:adhesion-energies}} and \textbf{Table~\ref{tbl:interface-free-energies}} is used.  The dashed and dotted lines indicate the equilibrium potentials for water reduction and oxidation, respectively.  At pH values above 10, \ce{SiO2} is known to dissolve in water, which is indicated by a gray hatching.~\cite{Pourbaix1974}}
  \label{fig:interface-Pourbaix-diagram}
  \vspace{-\baselineskip}
\end{figure*}

To determine which of the interface terminations of \textbf{Figure~\ref{fig:adhesion-energies}} are thermodynamically stable at conditions relevant for electrocatalysis, we applied the interface Pourbaix diagram formalism of section~\ref{sec:interface-formation-energy} to the interface structures of \textbf{Figure~\ref{fig:adhesion-energies}}.
\textbf{Table~\ref{tbl:interface-free-energies}} lists the expressions of the relative interface Gibbs free energies of formation for the different membrane terminations.
The free energies of the reconstructed and the hydrated interface show no pH or potential dependence, since they had no protons formally added or removed compared to the pristine \ce{SiO2}.
As noted above, the formal addition of a water molecule and simultaneous removal of a proton-electron pair is equivalent to the addition of a hydroxide ion, hence, \emph{hydroxylating} the cleaved interface is equivalent to \emph{deprotonating} the hydrated interface.

The resulting interface Pourbaix diagram is shown in \textbf{Figure~\ref{fig:interface-Pourbaix-diagram}a}.
Only three of the interface terminations occur in the phase diagram:
Irrespective of the pH value, the stable interfaces are (in order of increasing potential) the (1)~Si-terminated, (2)~reconstructed, and (3)~O-terminated membrane structures.
The hydrogenated, hydrated, and Hydroxylated membrane structures do not exhibit any stability region in the Pourbaix diagram.

The dashed and dotted lines in \textbf{Figure~\ref{fig:interface-Pourbaix-diagram}} indicate the boundaries of the electrochemical stability window of water under equilibrium conditions.
All three interface phases are accessible within the water stability range.
See also supporting \textbf{Figures S5--S7} for one-dimensional slices through the Pourbaix diagram for pH=0, pH=7, and $U$=0~V.
In absence of an applied potential, the \ce{SiO2} membrane is either Si terminated (for pH$<\sim$2) or reconstructed.

\ce{SiO2} is known to dissolve in water for pH values above 10 by forming \ce{HSiO3^-} or (for pH>12) \ce{SiO3^{2-}}~\cite{Pourbaix1974}, so that these pH regions are not accessible with an \ce{SiO2}-membrane coated Pt electrode.
Instead, the O-terminated membrane termination is only formed for positive applied potentials.

Note that the conditions for HER are at negative potentials below the equilibrium potential for water reduction (dashed line), for which the membrane is predicted to be generally Si-terminated in thermodynamic equilibrium.
At OER conditions, i.e., at potentials above the water oxidation potential (dotted line), the membrane is predicted to be O-terminated.

The equilibrium interface Pourbaix diagram of \textbf{Figure~\ref{fig:interface-Pourbaix-diagram}a} implies that water molecules can permeate through the \ce{SiO2} membranes instantaneously and without any kinetic barrier.
If the transport of water molecules towards and away from the buried interface is subject to a kinetic activation energy, the addition (removal) of water molecules to (from) the membrane surface would be kinetically hindered.
The formation of the reconstructed membrane surface is also likely associated with a kinetic barrier, since the structure of the reconstructed \ce{SiO2} surface is substantially different from the other membrane terminations (see also \textbf{Figure~\ref{fig:adhesion-energies}b}).
The Pourbaix diagram in \textbf{Figure~\ref{fig:interface-Pourbaix-diagram}b} shows the interface phases that are kinetically favored considering water transport and the kinetic limitation of the surface reconstruction, which are derived from the hydrated membrane at high positive potentials (shades of green/blue) and from the unreconstructed cleaved membrane at negative and low positive potentials (shades of brown).

The interface Pourbaix diagrams obtained from DFT calculations without van-der-Waals correction are shown in supporting \textbf{Figure~S8}.
As seen in the diagrams, the overall trends remain identical to the Pourbaix diagrams of \textbf{Figure~\ref{fig:interface-Pourbaix-diagram}}, but the size of the stability regions is affected and the hydrogenated membrane termination does not occur in the Pourbaix diagram with kinetically favored phases.
This provides further evidence of the importance of van-der-Waals interactions for the correct description of the membrane-catalyst interface.

\begin{figure*}
  \centering
  \includegraphics[width=\textwidth]{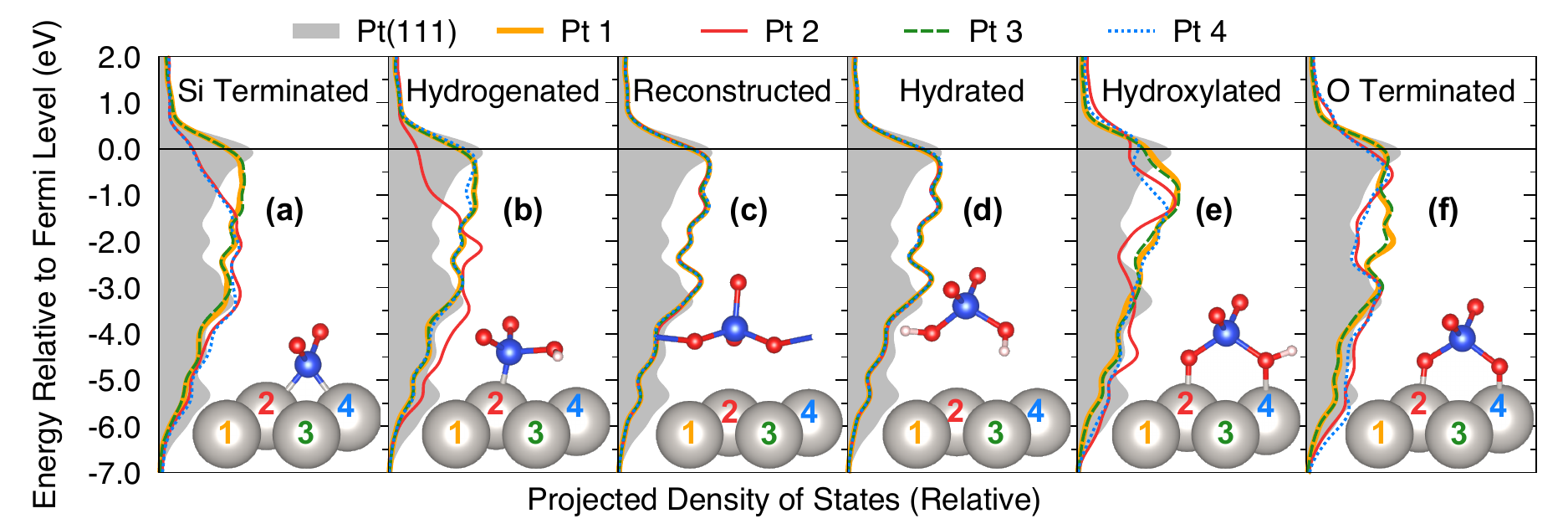}
  \caption{$d$-band projected electronic density of states (PDOS) of the four surface Pt atoms at the buried interfaces in contact with an \textbf{(a)}~Si-terminated, \textbf{(b)}~hydrogenated, \textbf{(c)}~reconstructed, \textbf{(d)}~hydrated, \textbf{(e)}~hydroxylated, and \textbf{(f)}~O-terminated \ce{SiO2} membrane.  The PDOS of a surface atom in the bare Pt(111) surface is shown for comparison (gray shaded).}
  \label{fig:PDOS}
\end{figure*}

\subsection{Charge transfer between the \texorpdfstring{\ce{SiO2}}{SiO2} membrane and the Pt surface}
\label{sec:charge-distribution}

The interface Pourbaix diagram of \textbf{Figure~\ref{fig:interface-Pourbaix-diagram}} shows that the stable phase of the buried \ce{SiO2 / Pt} interface depends on the pH value and the applied potential.
In turn, the amount of charge transfer between the \ce{SiO2} membrane and the \ce{Pt} surface can also be expected to vary with the prevalent bond at the buried interface.
Since the energy of the Pt $d$-band states is intimately connected with the catalytic activity ($d$-band model)~\cite{piss38-1991-103, n376-1995-238}, we evaluated the $d$-band projected electronic density of states (PDOS) of the surface Pt atoms for the different membrane terminations.

\textbf{Figure~\ref{fig:PDOS}} shows the PDOS of the surface Pt atoms for the different membrane terminations compared to the bare Pt surface, confirming that the \ce{SiO2} membrane has a significant impact on the electronic states of the Pt atoms at the buried interface.
In the case of the Si-terminated \ce{SiO2} membrane (\textbf{Figure~\ref{fig:PDOS}a}), the valence band DOS near the Fermi level is depressed for the two coordinated surface Pt atoms (Pt~2 and Pt~4), indicating partial oxidation, i.e., the Pt electrode is reducing the membrane.
The same is seen for the single coordinated Pt atom (Pt~2) at the hydrogenated interface (\textbf{Figure~\ref{fig:PDOS}b}).
The reconstructed and hydrated \ce{SiO2} membrane interacts only weakly with the Pt surface (see also \textbf{Figure~\ref{fig:adhesion-energies}}), showing no specific impact on any of the surface Pt atoms.
However, the valence band states are overall raised.
The DOS of the two coordinated Pt atoms (Pt~2 and 4) at the hydroxylated interface (\textbf{Figure~\ref{fig:PDOS}c}) show an increased valence-band density, indicating charge transfer from the membrane to the Pt surface, i.e., the Pt electrode is slighly oxidizing the membrane.
The impact on the DOS of the coordinated Pt atoms (Pt~2 and 4) is least obvious in the case of the O-terminated membrane (\textbf{Figure~\ref{fig:PDOS}d}).

\begin{figure}
  \centering
  \includegraphics[width=\colwidth]{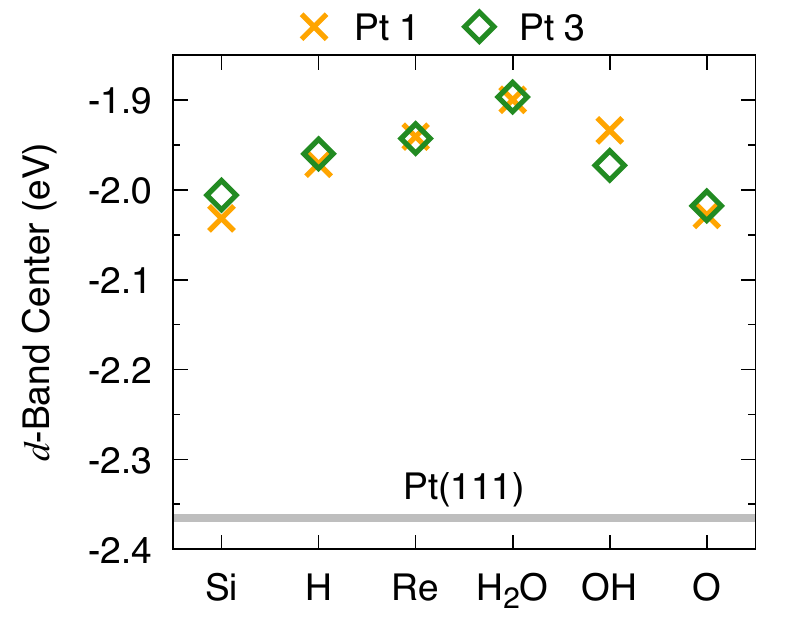}
  \caption{$d$-band centers of the two uncoordinated surface Pt atoms for each membrane termination.  \ce{Si} stands for \emph{Si-terminated}, \ce{H} for \emph{hydrogenated}, Re for \emph{reconstructed}, \ce{H2O} for \emph{hydrated}, \ce{OH} for \emph{hydroxylated}, and \ce{O} means \emph{O-terminated}.}
  \label{fig:dband}
\end{figure}

The Pt surface sites that are directly bonded with atoms in the \ce{SiO2} membrane might not be accessible for reactants, but the DOS of the vacant surface sites (Pt atoms 1 and 3 in \textbf{Figure~\ref{fig:PDOS}}) is also affected.
For all interface terminations, the valence-band DOS of the uncoordinated Pt atoms 3~and~4 increases near the Fermi level.
To quantify this change in the $d$-band DOS of the uncoordinated surface sites, we calculated the $d$-band center, and the results are compared to surface atoms in bare Pt(111) in \textbf{Figure~\ref{fig:dband}}.
As seen in the figure, the $d$-band center is raised by $\sim$0.4 to $\sim$0.5~eV.
The effect is strongest for the hydrogenated, hydrated, and hydroxylated interfaces, though the $d$-band center varies only by $\sim$0.1~eV across the four interface terminations.
Hence, reactants interacting with the Pt surface at the buried interface are experiencing transition-metal atoms with higher $d$-band center, which generally results in increased binding energies~\cite{potnaos108-2011-937, jpcc121-2017-20306}.

\begin{figure*}
  \centering
  \includegraphics[width=0.75\textwidth]{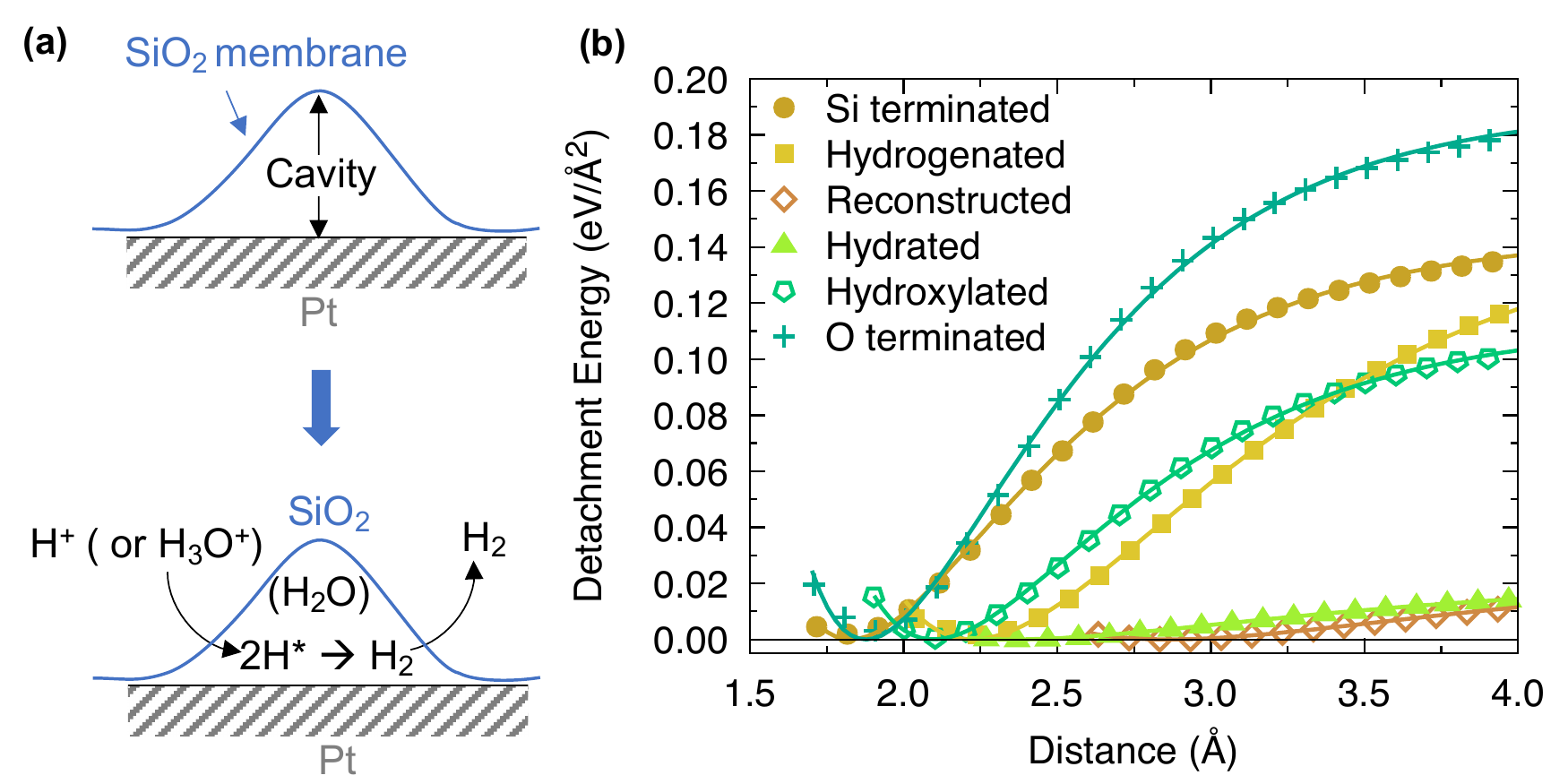}
  \caption{\textbf{(a)}~Schematic showing how the local detachment of the \ce{SiO2} membrane from the Pt surface leads to the formation of an interface cavity, and hydrogen evolution reaction taking place within an interface cavity. \textbf{(b)}~Energy curves for membrane detachment (\emph{detachment energy}) for the different membrane terminations.}
  \label{fig:cavity-formation}
\end{figure*}

\subsection{Towards determining active sites: cavities at the buried interface}
\label{sec:cavities}

Understanding mechanistically how electrochemical reactions take place at the buried \ce{SiO2 / Pt} interface would require modeling the interaction of reactants and products with the catalyst surface and the membrane and is thus beyond the scope of the present work.
However, a catalytic reaction can only occur at the buried \ce{SiO2 / Pt} interface if the reactants can access active sites at the surface of the Pt catalyst.
Hence, there should exist \emph{pockets} or \emph{cavities} at the buried interface that are sufficiently large for the reaction to take place.
Based on the interactions between the \ce{SiO2} membrane and the \ce{Pt} surface determined in the previous sections, we can derive initial insight into the formation of such interface cavities.

The formation of an interface cavity necessarily involves the local detachment of the \ce{SiO2} membrane from the Pt surface, a process that is schematically shown in \textbf{Figure~\ref{fig:cavity-formation}a}.
As an approximate measure of the energy required for cavity formation, we determined the detachment energy as defined in equation~\eqref{eq:detachment-energy} for membranes with the four stable terminations.

The energies required for complete detachment are equivalent to the adhesion energies of \textbf{Table~\ref{tbl:adhesion-energies}}, however, complete detachment is not needed to form an interface cavity.
The space requirement for the catalytic reaction depends instead on the reactant and product species.
Considering the small size of \ce{H2} molecules with a bond length of only 0.74~\AA{}, cavities with a diameter of 2~\AA{} could be sufficient for HER.
The diameter of a water molecule is about 2.75~\AA{}, so that more space is needed if the reactants or products are dissolved.

\textbf{Figure~\ref{fig:cavity-formation}b} shows the detachment energies of membranes with the four different stable terminations as function of the distance from the Pt surface.
As seen in the figure, the curvature of the bond curves varies strongly with the membrane termination:
The O- and Si-terminated membranes, which bind most strongly to the Pt surface (\textbf{Table~\ref{tbl:adhesion-energies}}), also exhibit the steepest increase in the detachment energy, indicating a stiff bond that might impede the formation of a cavity at the interface.
The hydrated and hydrogenated membrane terminations show softer bonds with the Pt surface, allowing the more facile creation of interface cavities.
The reconstructed and hydrated membranes bind only via weak interactions with the Pt surface and show the smallest detachment energies.

To give a concrete example, creating a cavity with 3~\AA{} diameter at the O-terminated interface approximately requires an energy of about 0.14~eV/\AA{}$^{2}$.
Even though, the adhesion energy of Si-terminated and hydrogenated membranes is nearly identical (\textbf{Table~\ref{tbl:adhesion-energies}}), the distance of the membranes from the Pt surface and the curvature of the detachment energy are quite different.
While Si-terminated membranes bind stiffly with the Pt surface at a distance of around 1.8~\AA{}, the bonding with hydrogenated membranes is softer, and only 0.06~eV/\AA{}$^{2}$ are required to create a 3~\AA{} cavity.
Nearly no energy is required to create 3~\AA{} cavities at the reconstructed and hydrated interfaces.

\section{Discussion}
\label{sec:discussion}

In this article, we developed a formalism for the description of semipermeable coatings on electrocatalysts in aqueous electrolyte.
We evaluated the impact of the pH value and an applied potential on the interactions between the Pt catalyst surface and a coated \ce{SiO2} membrane by extending the concept of Pourbaix diagrams to (permeable) interfaces.
The results show that the structure and composition of the buried \ce{SiO2 / Pt} interface are not static but instead respond to changes in the pH value and the applied potential.
Our calculations predict (\textbf{Figure~\ref{fig:interface-Pourbaix-diagram}a}) that the \ce{SiO2} membrane in contact with the Pt surface is Si-terminated for potentials below the water reduction potential.
As the potential is increased, the \ce{SiO2} membrane preferentially exhibits a reconstruction that is also known to be stable in vacuum and binds only with dispersive interactions to the Pt surface.
As the potential for water oxidation is approached, the membrane is preferentially O-terminated, i.e., \ce{Si-O^-} groups bind to the Pt surface.

Considering kinetic barriers for water transport through the \ce{SiO2} membrane and for the formation of the reconstructed \ce{SiO2} surface, other interface phases might become accessible (\textbf{Figure~\ref{fig:interface-Pourbaix-diagram}b}).
In the order of increasing potential, the membrane is Si-terminated at negative potentials, then the addition of a proton to the pristine membrane surface (hydrogenation) becomes thermodynamically preferred, so that silanol (\ce{Si-OH}) groups are formed, then the preferred membrane termination becomes hydroxylated, i.e., exhibits silanol and \ce{Si-O^-} groups before eventually reaching the O-terminated phase.

The trends predicted by our surface Pourbaix diagram analysis agree largely with chemical intuition.
The formation and dissociation of silanol groups \ce{Si-O^- + H2O <-> Si-OH + OH^-} on the surface of \ce{SiO2} in aqueous solution has been well characterized, and silanol groups are known to be present for pH$\lesssim$4, i.e., at pH values smaller than the p$K_{a}$ of \ce{Si-OH}~\cite{cpl191-1992-327,jpcl3-2012-1269}.
The phase dynamics at the buried interface is, however, more complex than the surface phases of \ce{SiO2} alone, since dangling bonds on the membrane surface can be saturated by coordinating with the Pt surface atoms.
Cyclic voltammetry characterization of the \ce{SiO2 / Pt} system has shown evidence of the formation of \ce{Pt-O} bonds at positive potentials~\cite{ac8-2018-11423}, in agreement with our predictions.
An experimental confirmation of the existence of \ce{Si-Pt} bonds at negative voltages should in principle be possible with vibrational spectroscopy, but we are not aware of any published reports that we could compare with.

We note that Pt metal reacts with water at positive potentials above 1.18~V~vs.~SHE at pH~=~0 to form the oxide \ce{PtO2}~\cite{Pourbaix1974}.
\ce{PtO2} formation was not considered in our calculations and might occur also at the buried \ce{SiO2 / Pt} interface.
However, at the relevant potentials the formation of \ce{PtO2} competes with the \ce{Pt-O} bonds that are present at the O-terminated interface, so that the \ce{SiO2} membrane might have a stabilizing effect on the Pt surface.
Further investigation is needed to determine whether Pt oxide formation is likely.

The dynamic changes of the buried interface with pH and potential can be expected to affect catalytic performance.
First, the bonding and adhesion of the membrane to the Pt surface controls not only the stability against unwanted detachment but also the accessible catalyst surface area.
In an electrocatalytic reaction, the reactants have to be able to reach the surface of the Pt catalyst, which is not possible if the \ce{SiO2} membrane binds too tightly to the Pt surface.
At pH values and potentials for which the membrane binds strongly with the Pt surface, i.e., with \ce{O-Pt} bonds (\textbf{Figure~\ref{fig:adhesion-energies}}) at positive potentials, the creation of interface cavities near the catalyst surface requires more energy and is therefore less likely, thus resulting in a decrease of the accessible surface area.
The stiffness of the membrane--catalyst bond additionally determines the flexibility of the buried interface with respect to partial detachment of the \ce{SiO2} membrane
(\textbf{Figure~\ref{fig:cavity-formation}}).

Our calculations predict that cavities at the interface are most likely to form when the \ce{SiO2} membrane is either reconstructed or hydrated, i.e., at zero or small potentials and within the thermodynamic stability region of water.
We note that the detachment energy is only one factor for the formation of interface cavities, and the mechanical response of the \ce{SiO2} membrane upon deformation and, potentially, the interaction with water within the cavities are also needed for a complete picture.
If \ce{SiO2} membranes are permeable for water, which has not been conclusively determined, once formed, interface cavities might be stabilized by water at the interface.
While water is implicitly considered in our formalism, a future study should investigate the impact of explicit water molecules at the buried interface.

Not only the presence but also the size of interface cavities will affect the mechanism of catalytic reactions at the buried Pt surface.
Exploiting nanoscale confinement effects has previously been proposed for the reduction of overpotentials for oxygen evolution reaction over \ce{RuO2}~\cite{c7-2015-738}.
The pH and potential dependence of the membrane detachment energy, together with temperature, might provide control knobs for the optimization of cavity sizes that should be further investigated.

In addition to confinement effects, the \ce{SiO2} membrane also affects the electronic structure of the Pt catalyst, as our analysis of the $d$-band DOS exemplifies (\textbf{Figure~\ref{fig:PDOS}}).
The surface Pt atoms that are directly coordinated by membrane atoms are partially oxidized at low potentials, i.e., the membrane is reduced in contact with the Pt electrode consistent with expectations.
This charge transfer becomes smaller as the potential increases towards the water oxidation potential.
However, irrespective of the membrane termination and potential bias, the $d$-band states of the uncoordinated surface Pt sites that are accessible to reactant species is raised to higher energies compared to the bare Pt(111) surface.

An increase of the $d$-band center by $\sim$0.4~eV has previously been shown to lead to a comparable increase in the adsorption energy of small molecules~\cite{potnaos108-2011-937}, which could have a significant impact on the catalytic activity.
Considering the volcano plot for HER activity~\cite{nm5-2006-909, ees6-2013-1509}, an increased hydrogen binding energy could take metals that bind hydrogen more weakly than Pt closer to the Sabatier optimum.
The \ce{SiO2} membrane might therefore play a more important role for the catalytic activity of metals that are less efficient HER catalysts than Pt.

Further differences in the interaction of \ce{SiO2} membranes with other transition metals can be expected: depending on the electronegativity (or nobility) of the metal, we expect the interface to be more or less oxidized compared to Pt.
For example, a less noble metal such as Cu can be expected to exhibit a stronger preference for oxygen at the interface, i.e., the phase boundary of the Si-terminated membrane in the SiO2/Cu interface Pourbaix diagram should shift to lower pH values and more negative potentials compared to SiO2/Pt.
No such differences would be expected for Rh, which has the same Pauling electronegativity as Pt (2.28).
Indeed, the adhesion energy of the Si-terminated membrane is lower on \ce{Cu(111)} than on \ce{Pt(111)} but nearly identical on \ce{Rh(111)} (\textbf{Table S2}), though the construction of the complete interface Pourbaix diagram will be required to make quantitative statements.

Finally, our approach is based on a number of assumptions that should be considered when comparing with experiment.
The calculated surface Pourbaix diagram reflects thermodynamic equilibrium conditions, though it can be expected that there is also a kinetic barrier associated with the transition between different bonds and membrane terminations.
One contribution to such barrier is the detachment energy of \textbf{Figure~\ref{fig:cavity-formation}}, which indicates that the stiffness of the \ce{Si-Pt} bond would give rise to an activation energy for the addition of water molecules or \ce{OH} groups to the membrane surface.
Hence, the bonding at the interface is unlikely to change instantaneously with pH and potential adjustments.
Additionally, our analysis is based on a homogeneous crystalline \ce{SiO2} membrane structure, whereas the coating of actual membrane-coated electrocatalysts is amorphous and may exhibit off-stoichiometries.
We expect real-world \ce{SiO2} membranes to exhibit a distribution of different bonds with the Pt surface instead of the homogeneous bonding modeled here, but the bond distribution is expected to center around the equilibrium bond predicted by the interface Pourbaix diagram.
The direct modeling of amorphous interface structures on the atomic scale, for example using machine-learning techniques~\cite{jpe1-2019-32002, acsc10-2020-9438}, is currently an active field of research.
However, amorphous \ce{SiO2} exhibits the same \ce{SiO4} building blocks as the crystalline structure modeled here, so that the chemical properties remain largely unchanged, and any interface bond distributions should therefore be dominated by the thermodynamically most stable bonds identified in the present work.

\section{Conclusions}
\label{sec:conclusions}

Semipermeable silica coatings are promising for the protection of transition-metal electrocatalysts from corrosion and from poisoning by contaminants.
We investigated how \ce{SiO2} membranes bind to Pt catalyst surfaces and how this interaction affects properties relevant for catalysis.
By extending the concept of Pourbaix diagrams to electrochemical solid-solid interfaces of proton-permeable membranes, we determined the stable termination of the \ce{SiO2} membrane at the buried interface as a function of the pH value of the electrolyte and the applied potential.
Our calculations predict that the thermodynamically preferred membrane termination varies between the potentials for water reduction and oxidation, and the terminating membrane species changes from \ce{Si} to \ce{Si-O} as the potential increases.
At intermediate potentials, the \ce{SiO2} membrane is terminated by a reconstruction and binds only weakly to the Pt surface, unless the formation of the reconstruction is kinetically hindered, in which case \ce{Si-OH} and \ce{SiO-OH} groups will bind to the Pt surface.
This variation of the membrane termination affects the catalytic properties of the \ce{SiO2 / Pt} system twofold: (i)~through differences in the charge transfer from the Pt surface to the \ce{SiO2} membrane that alter the position of the Pt $d$~band, and (ii)~through changes in the strength and stiffness of the membrane-Pt bonding that control the formation of cavities at the buried interface and the Pt surface area that is accessible by reactant species.
Charge transfer between the Pt surface and the \ce{SiO2} membrane leads to a general increase in the $d$-band center of the uncoordinated Pt surface atoms, which is expected to increase the binding energy of small molecules compared to the bare \ce{Pt(111)} surface.
\ce{SiO2} membranes are therefore not simply inactive protective coatings but instead participate actively in the electrocatalytic reaction, and the parameters of the combined membrane-coated electrocatalyst are available for tuning the catalytic activity and selectivity.
Future studies might investigate how the synergy of an oxide membrane and a metal catalyst can be exploited for the design of better catalysts for specific electrocatalytic processes, such as seawater electrolysis.


\section*{Acknowledgments}

This work has been partially supported by the Alfred P.~Sloan Foundation
through grant number G-2020-12650.
Computing resources from Columbia University's Shared Research Computing
Facility project are acknowledged, which is supported by NIH Research
Facility Improvement Grant 1G20RR030893-01, and associated funds from
the New York State Empire State Development, Division of Science
Technology and Innovation (NY STAR) Contract C090171, both awarded April
15, 2010.
The authors thank Daniel V.~Esposito, Marissa E.~S.~Beatty, and Nongnuch
Artrith for insightful discussions.


\clearpage\newpage
\onecolumngrid
\appendix
\renewcommand{\thefigure}{S\arabic{figure}}
\renewcommand{\thetable}{S\arabic{table}}
\setcounter{figure}{0}
\setcounter{table}{0}

\section{Supporting Information}

\section*{S1. Supporting Tables}

\begin{table}[H]
  \centering
  \caption{Comparison of the calculated lattice parameters of
    face-centered cubic \ce{Pt} and $\alpha$-quartz \ce{SiO2} with
    reference values from the literature.}
  \label{tbl:lattice-parameters}
  \begin{tabular}{lllc}
    \toprule
              & \textbf{DFT (This Work)} & \textbf{Experiment} & \textbf{Reference} \\
    \midrule
    \ce{Pt}   & $a=b=c=3.97$~\AA{} & $a=b=c=3.92$~\AA{} & [\citenum{ea50-2005-5384}] \\
              & $\alpha=\beta=\gamma=90^{\circ}$ & $\alpha=\beta=\gamma=90^{\circ}$  \\
    \ce{SiO2} & $a=b=5.02$~\AA{}, $c=5.51$~\AA{} & $a=b=4.91$~\AA{}, $c=5.41$~\AA{} & [\citenum{ssc72-1989-507}] \\
              & $\alpha=\beta=90^{\circ}$, $\gamma=60^{\circ}$ & $\alpha=\beta=90^{\circ}$, $\gamma=60^{\circ}$   \\
    \bottomrule
  \end{tabular}
  \vspace{-0.5\baselineskip}
\end{table}

\begin{table}[H]
  \centering
  \caption{Calculated adhesion energies of Si-terminated \ce{SiO2} membranes on different metal substrates.}
  \label{tbl:adhesion-energies-Pt-Cu-Rh}
  \begin{tabular}{cc}
    \toprule
    \textbf{Metal substrate} & \textbf{Adhesion energy (eV/\AA{}$^{2}$)} \\
    \midrule
\ce{SiO2 / Cu(111)} & $-0.098$ \\
\ce{SiO2 / Pt(111)} & $-0.146$ \\
\ce{SiO2 / Rh(111)} & $-0.143$ \\
    \bottomrule
  \end{tabular}
  \vspace{-0.5\baselineskip}
\end{table}

\section*{S2. Supporting Figures}

\begin{figure}[H]
	\centering
	\includegraphics[width=0.8\textwidth]{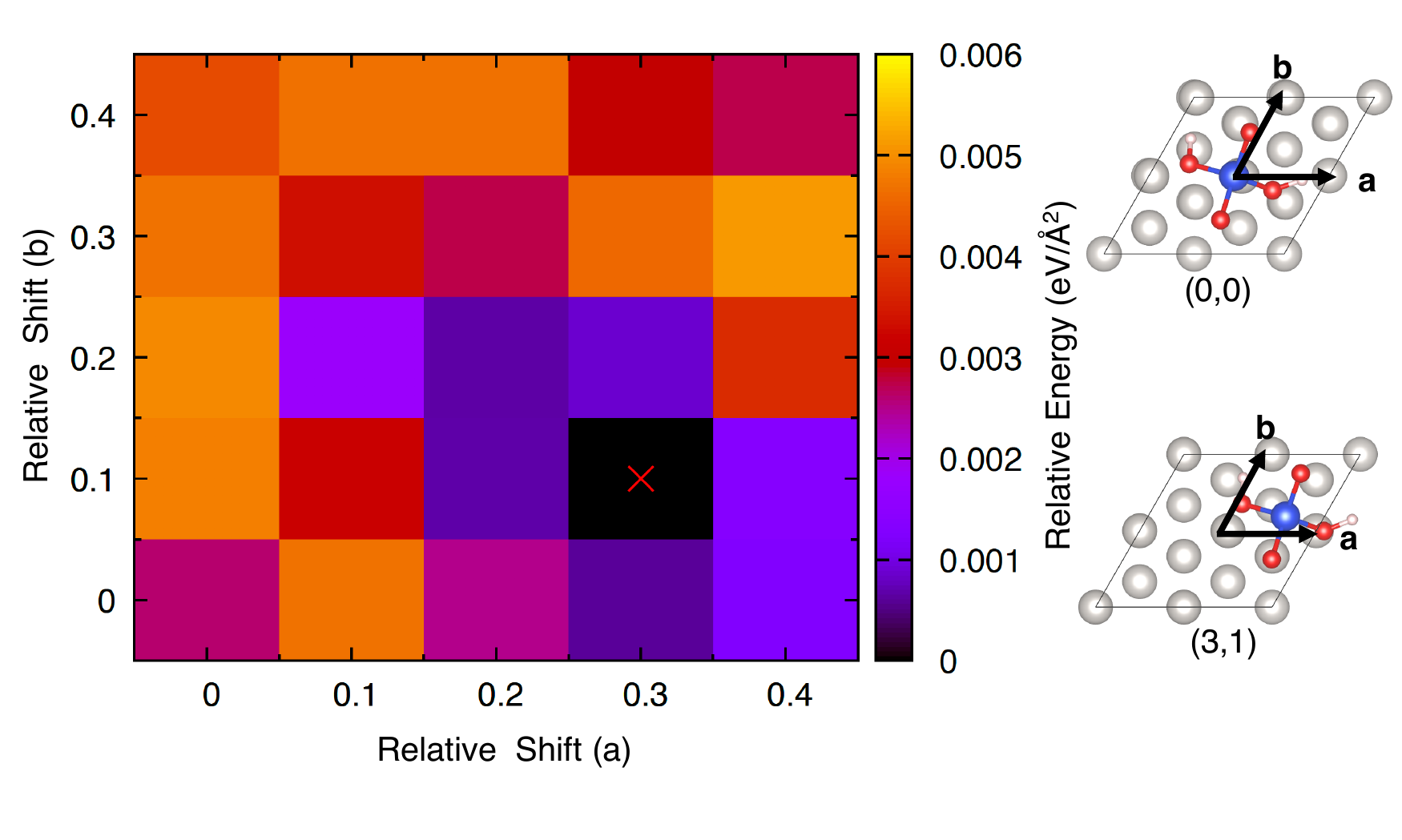}
	\caption{Relative energy of the hydrated membrane at different binding sites on the Pt(111) surface. The binding site is referenced to the center of the Si atom closest to the Pt surface, as shown in the figures.}
	\label{SI-fig:binding-positions-hydrated}
\end{figure}

\begin{figure}[H]
	\centering
	\includegraphics[width=0.9\textwidth]{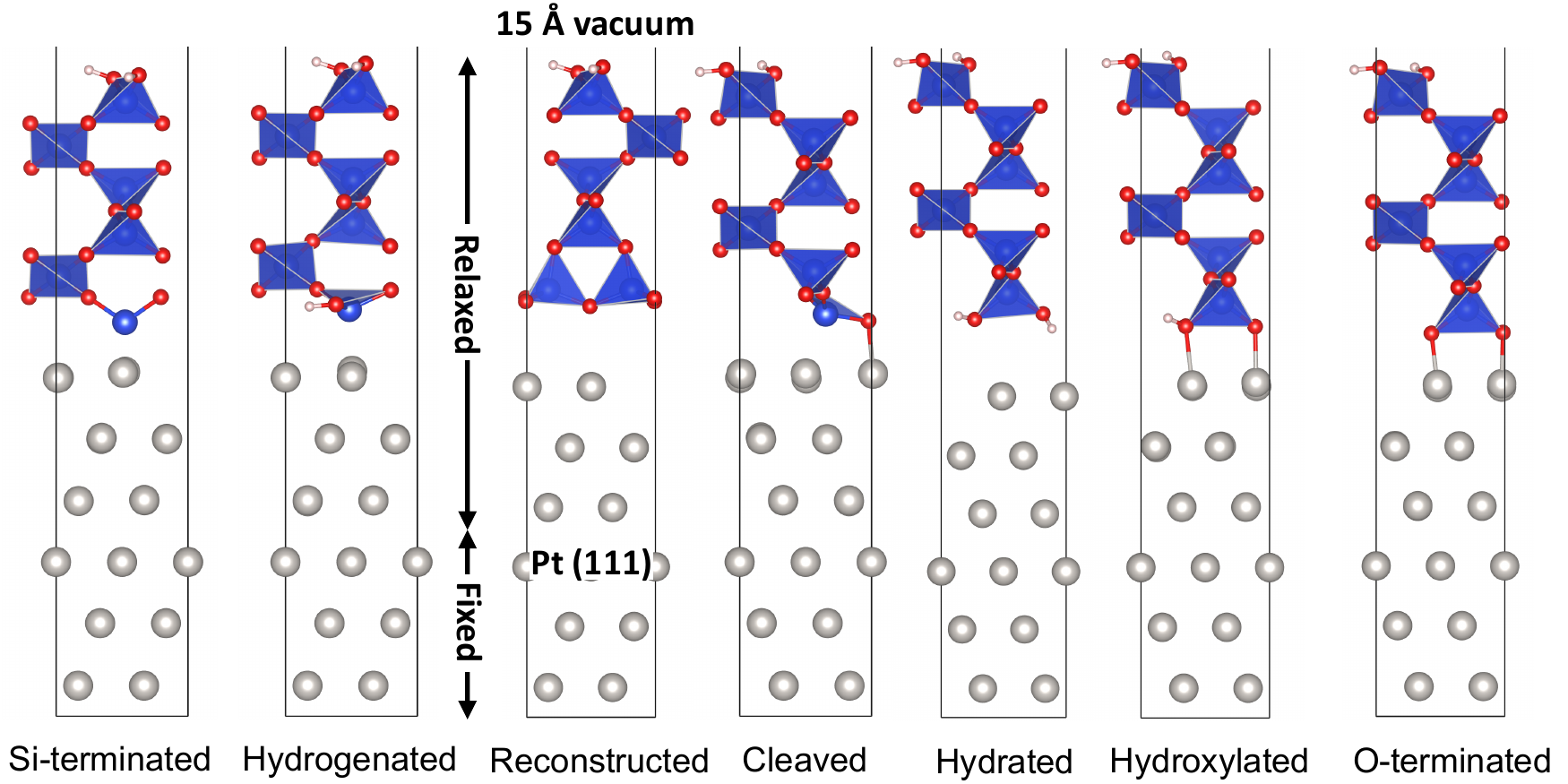}
	\caption{Visualization of the interface structure models for different membrane terminations. \ce{SiO4} tetrahedra (blue) are highlighted to indicate the fundamental building block of silica.}
	\label{SI-fig:interface-structures}
\end{figure}

\begin{figure}[H]
	\centering
	\includegraphics[width=0.9\textwidth]{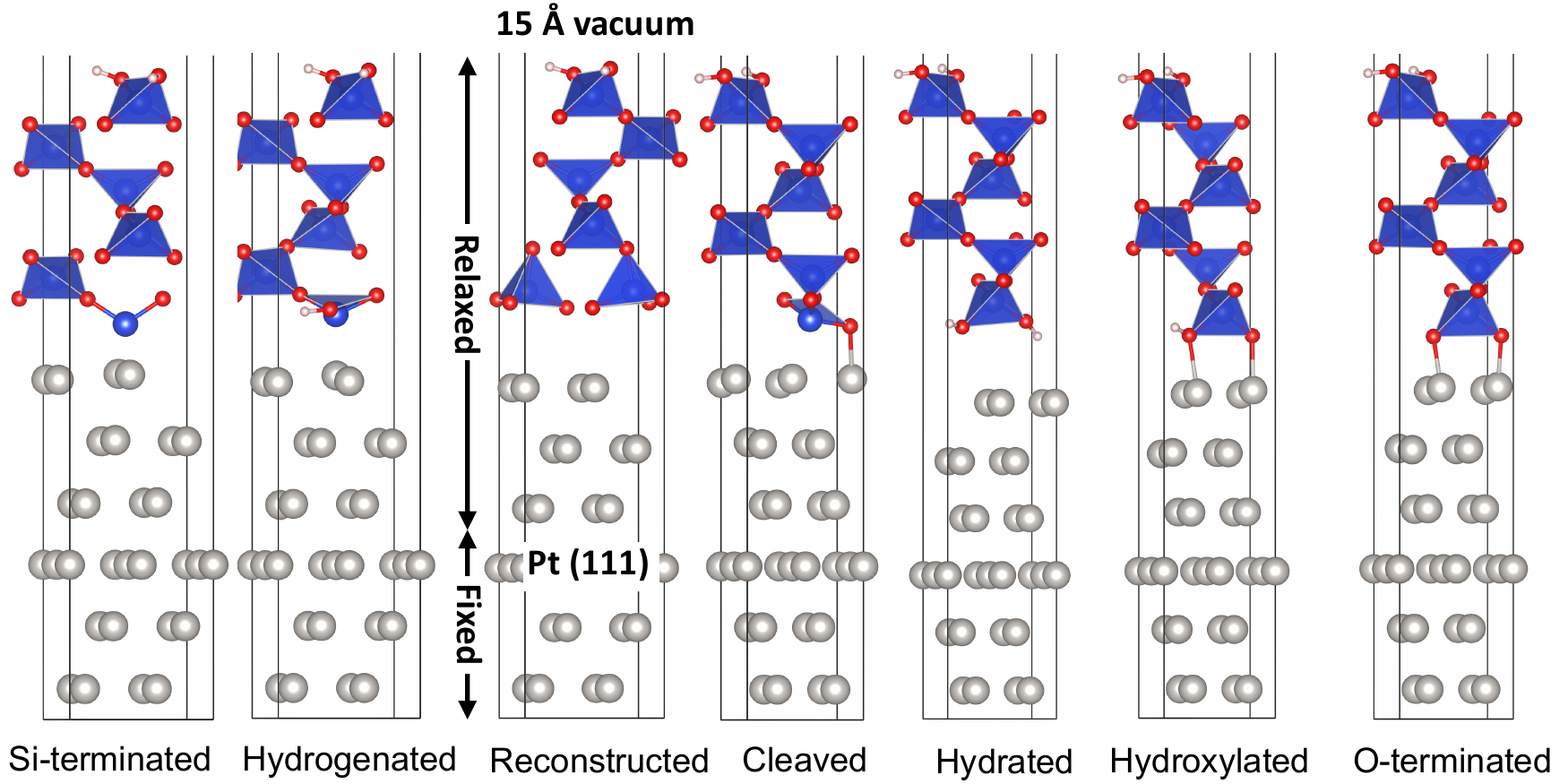}
	\caption{The structures of \textbf{Figure S8} but rotated by 10~$^{\circ}$ around the third lattice vector.  Structure files with atomic coordinates are also provided as additional supporting information.}
	\label{SI-fig:interface-structures-rotation}
\end{figure}

\begin{figure}[H]
	\centering
	\includegraphics[width=0.9\textwidth]{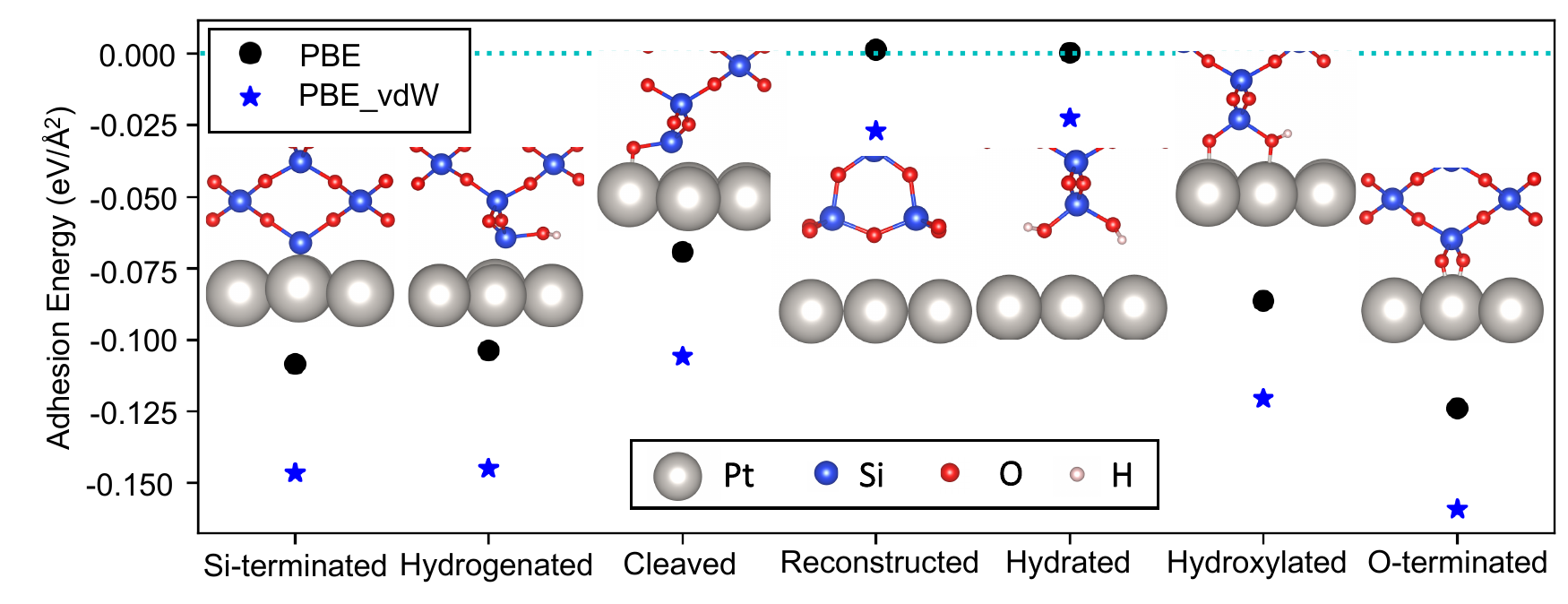}
	\caption{Adhesion energies for different membrane terminations as obtained from uncorrected DFT calculations (black point) and including the D3 van-der-Waals correction (blue star). Shown as insets are the corresponding atomic structures.}
	\label{SI-fig:adhesion-energy-vdW}
\end{figure}

\begin{figure}[H]
    \centering
    \includegraphics[width=0.8\textwidth]{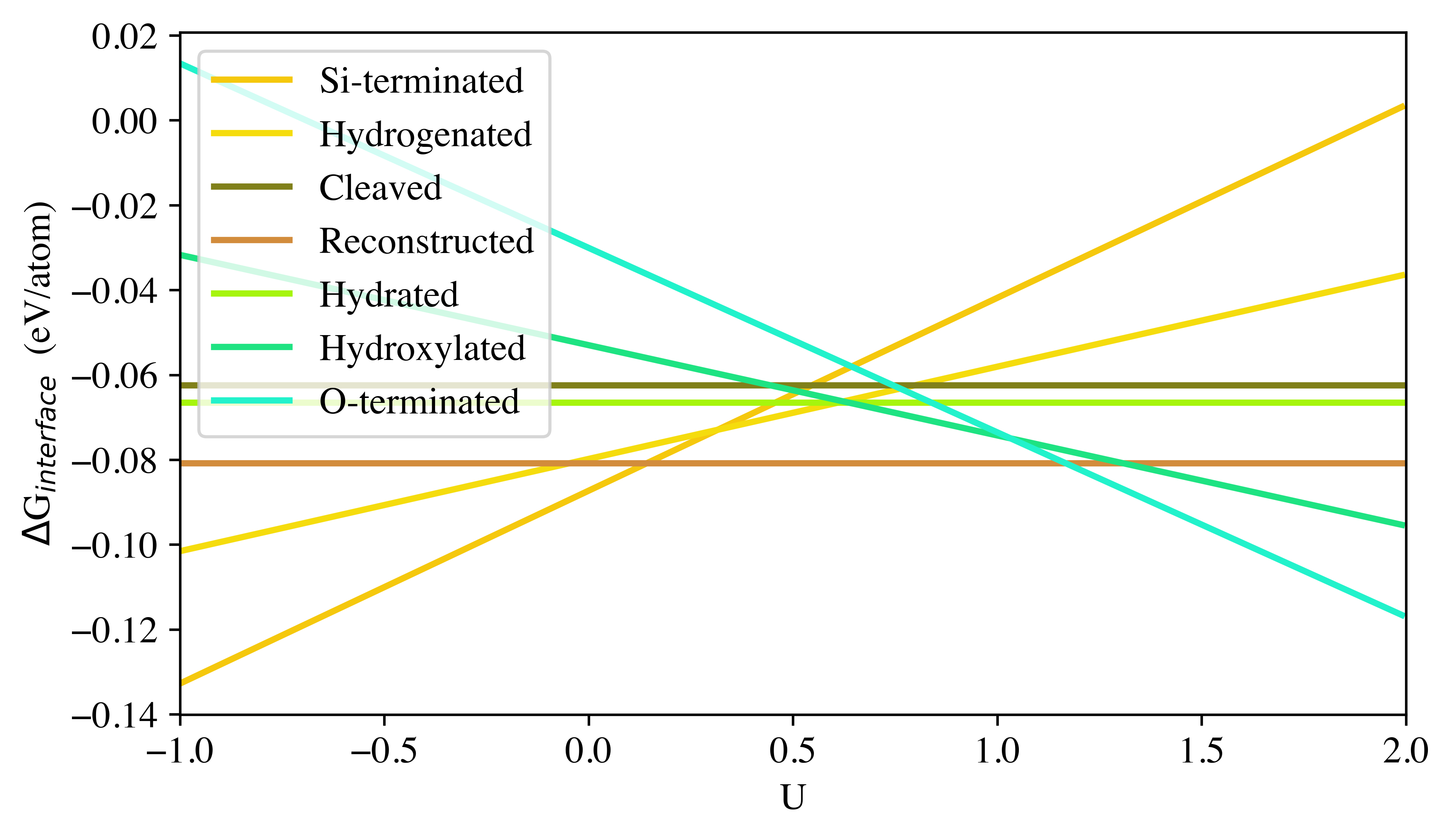}
    \caption{Gibbs formation energies of buried interfaces with different membrane terminations \textbf{at pH = 0} as function of the applied potential.}
    \label{SI-fig:Gibbs-vdW-pH=0}
\end{figure}

\begin{figure}[H]
	\centering
	\includegraphics[width=0.8\textwidth]{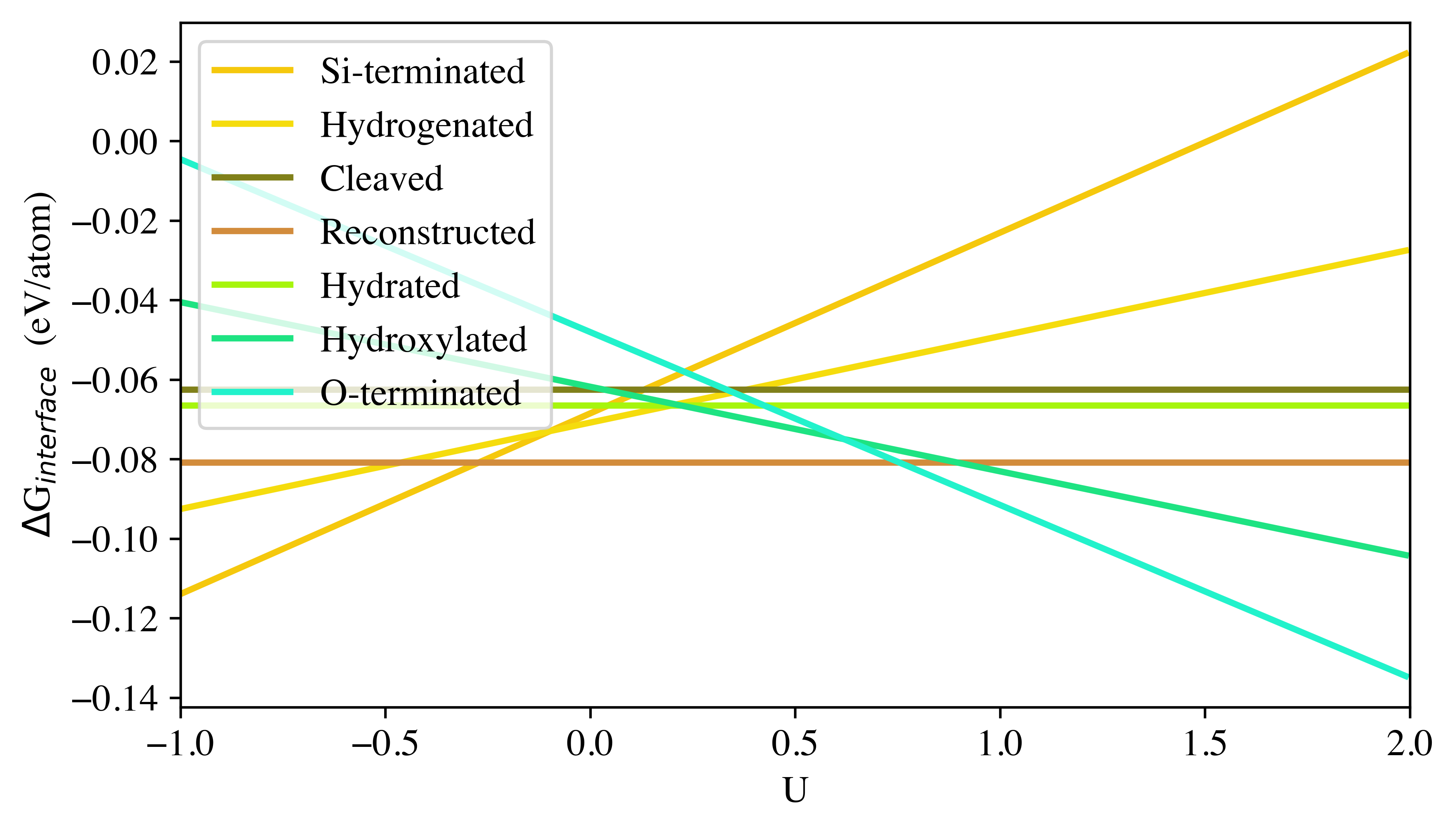}
	\caption{Gibbs formation energies of buried interfaces with different membrane terminations \textbf{at pH = 7} as function of the applied potential.}
	\label{SI-fig:Gibbs-vdW-pH=7}
\end{figure}

\begin{figure}[H]
	\centering
	\includegraphics[width=0.8\textwidth]{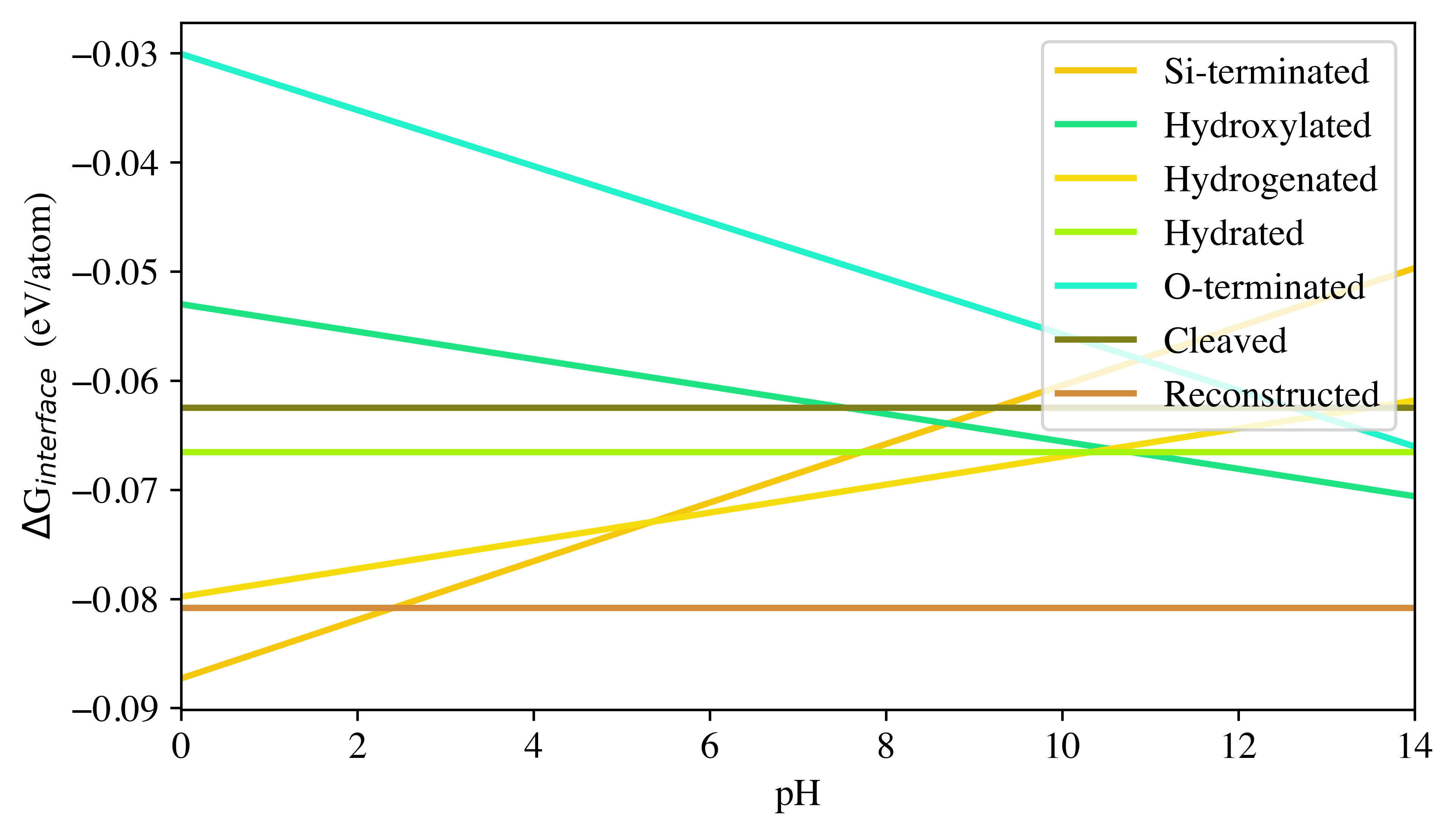}
	\caption{Gibbs formation energies of buried interfaces with different membrane terminations \textbf{at $U$ = 0~V} as function of the pH value.}
	\label{SI-fig:Gibbs-vdW-U=0}
\end{figure}

\begin{figure}[H]
	\centering
    \includegraphics[width=\textwidth]{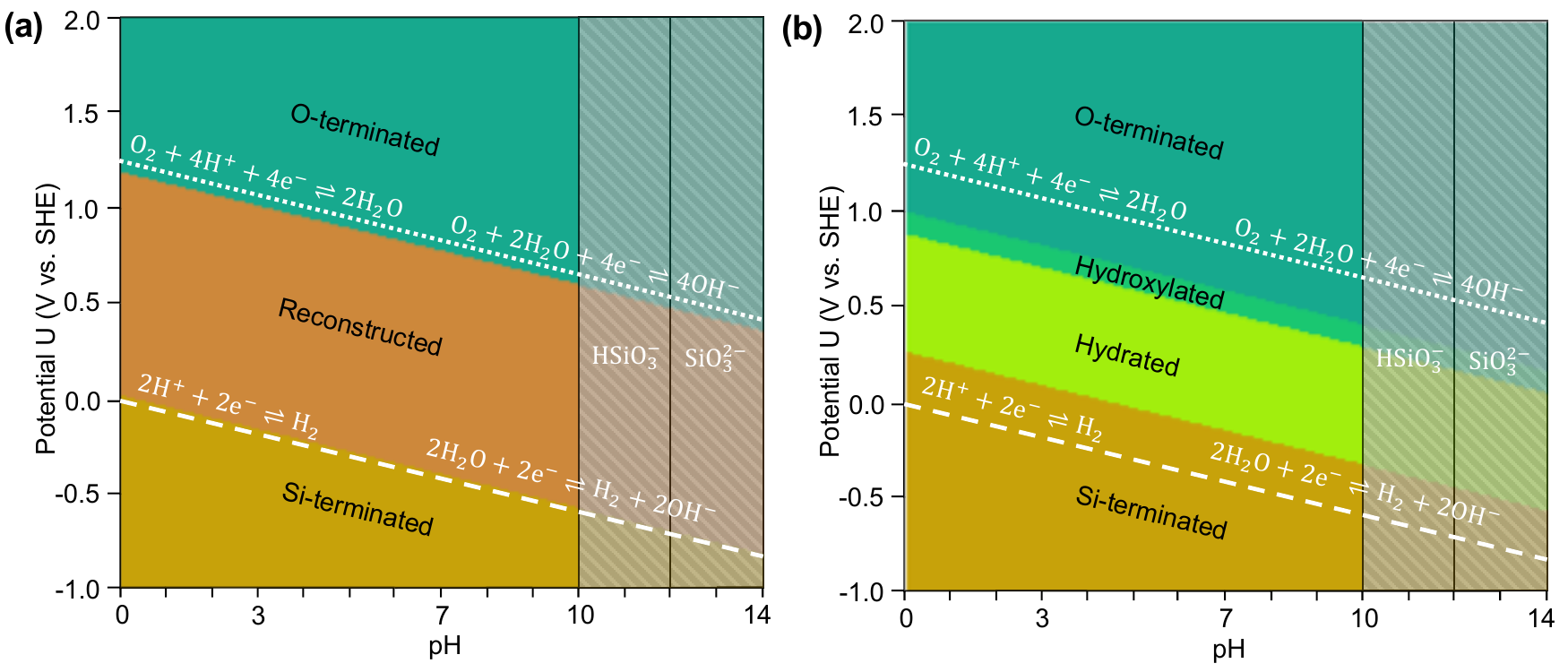}
	\caption{Calculated interface Pourbaix diagrams without van-der-Waals correction. \textbf{(a)}~assuming no transport limitation for water through the \ce{SiO2} membrane and \textbf{(b)}~with kinetically stabilized regions due to transport limitation for water. Different colors indicate the stability regions of different interface terminations. The dashed and dotted lines indicate the equilibrium potentials for water reduction and oxidation, respectively.  At pH values above 10, \ce{SiO2} is known to dissolve in water, which is indicated by a gray hatching.~\cite{Pourbaix1974}}
	\label{SI-fig:Pourbaix-diagram}
\end{figure}

\section*{S3. Contributions to the interface free energy}

The interface Pourbaix diagram formalism introduced in the main manuscript describes the dependence of the interface free energy on the pH value and the electrode potential including the temperature dependence of the standard hydrogen electrode, but it neglects the differences in temperature dependence of the free energy of the solid and liquid phases.
Here, we provide a justification for this approximation.

The only liquid phase involved in our calculations is liquid water.
However, the term due to the free energy of water is identical in all interface Gibbs free energies of formation, so that any error made in this contribution will only result in a constant energy shift but will not affect any relative quantities that compare different interfaces, such as the Pourbaix diagram.
The same argument can be made for the Pt catalyst, which is identical in all interface models.
The only component that varies across the interface models is the \ce{SiO2} membrane.

\textbf{Table~2} in the main manuscript gives the Gibbs free energy of interface formation as functions of the pH value and the electrode potential $U$.
As seen in the table, the Gibbs energies of four of the interface compositions varies with at least $1\times{}U$, corresponding to a variation of at least 3~eV for the considered potential range from -1~V to +2~V.
The pH dependence is $\sim{}20$~times smaller per pH unit and gives rise to a variation of 0.8~to~1.7~eV for the considered pH range (0~to~14).

To investigate the temperature dependence of free energies, we  derive the free energy change when increasing the temperature from 0~K to 298.15~K (room temperature)
\begin{align}
    G^{298.15 K}_{A} - G^{0K}_{A}
    = \Delta G_{A}
    = \Delta H - \Delta (TS)
    = \Delta U + \Delta (PV) - \Delta(TS)
    \label{eq:thermo-definition}
\end{align}
where $\Delta H$ and $\Delta U$ are the change of enthalpy and internal energy, and $\Delta \left( PV \right)$ and $\Delta \left( TS \right)$ are the energy terms of volume change and entropy change, respectively.
Assuming constant pressure $P$ in this temperature range yields
\begin{align}
    \Delta G_{A}
    = \Delta U + P \cdot \Delta V - \Delta (TS)
    \approx P \cdot \bigl(V^{298K} - V^{0K}\bigr) - \Delta (TS)
    \label{eq:thermo-approx}
\end{align}
where the specific internal energy change $\Delta U$ from $0K$ to $298.15K$ is neglected.
Thus the temperature dependence of the free energy mainly originates from two terms:
the volume work $P \cdot (V^{298K} - V^{0K})$ and the increase of entropy $\Delta (TS)$.

At room temperature, the density of quartz \ce{SiO2} is $2.65 g/cm^3$~\cite{aca175-1985-99}, corresponding to a molar volume of $22.64 \times 10^{-6} m^3/mol$.
The volumetric thermal expansion of quartz \ce{SiO2} at room temperature compared to 0~K is approximately 0.6\% \cite{joap53-1982-6751}, which corresponds to a volume work of 0.02~J/mol, i.e., a tiny energy difference compared to the pH and potential dependence.

The entropy term is given by the integral $\Delta (TS) = \int_{0}^{298.15} S \, \delta T$, which is 6.36~KJ/mol for \ce{SiO2}~\cite{jpcb105-2001-6025}.
Hence, the free energy decrease of \ce{SiO2} owing to the $TS$ term is $-6.36 KJ/mol = -0.066 eV/atom$ at room temperature, a much larger contribution than the $PV$ term.
However, since the various interface structures considered in the main manuscript are mostly identical and differ only by at most 4 atoms in the \ce{SiO2 / Pt} interface region, the energy differences arising from the entropy term are significantly smaller than the free energy differences due to pH and potential dependence.

\twocolumngrid

\providecommand{\latin}[1]{#1}
\makeatletter
\providecommand{\doi}
  {\begingroup\let\do\@makeother\dospecials
  \catcode`\{=1 \catcode`\}=2 \doi@aux}
\providecommand{\doi@aux}[1]{\endgroup\texttt{#1}}
\makeatother
\providecommand*\mcitethebibliography{\thebibliography}
\csname @ifundefined\endcsname{endmcitethebibliography}
  {\let\endmcitethebibliography\endthebibliography}{}

\end{document}